\documentclass{aastex631}

\usepackage[T1]{fontenc}
\usepackage{graphicx}	
\usepackage{amsmath}	
\usepackage{amssymb}	
\usepackage{wrapfig}
\usepackage{natbib}
\usepackage{xcolor}
\usepackage{ulem}
\usepackage{soul}
\usepackage{hyperref}

\usepackage{nopageno}
\usepackage{geometry}
\usepackage{xcolor}
\definecolor{blazeorange}{rgb}{1.0, 0.4, 0.0}
\definecolor{seagreen}{rgb}{0.18, 0.55, 0.34}
\definecolor{rufous}{rgb}{0.66, 0.11, 0.03}
\definecolor{royalfuchsia}{rgb}{0.79, 0.17, 0.57}
\definecolor{scarlet}{rgb}{1.0, 0.13, 0.0}
\definecolor{royalpurple}{rgb}{0.47, 0.32, 0.66}
\definecolor{darkblue}{rgb}{0, 0, 0.66}

\DeclareRobustCommand{\VAN}[3]{#2}
\let\VANthebibliography\thebibliography
\def\thebibliography{\DeclareRobustCommand{\VAN}[3]{##3}\VANthebibliography}

\hypersetup{linkcolor=blue,citecolor=blue,filecolor=cyan,urlcolor=magenta}
\usepackage{graphicx}   
\usepackage{amsmath}    
\usepackage{amssymb}    
\usepackage{newtxtext,newtxmath}
\usepackage{color}
\usepackage{threeparttable}
\usepackage{hyperref}
\usepackage{ulem}

\def\d{\mathrm{d}}
\newcommand{\lta}{\lower 2pt \hbox{$\, \buildrel {\scriptstyle <}\over {\scriptstyle \sim}\,$}}
\newcommand{\gta}{\lower 2pt \hbox{$\, \buildrel {\scriptstyle >}\over {\scriptstyle \sim}\,$}}
\definecolor{blazeorange}{rgb}{1.0, 0.4, 0.0}
\definecolor{seagreen}{rgb}{0.18, 0.55, 0.34}
\definecolor{rufous}{rgb}{0.66, 0.11, 0.03}
\definecolor{royalfuchsia}{rgb}{0.79, 0.17, 0.57}
\definecolor{scarlet}{rgb}{1.0, 0.13, 0.0}
\definecolor{royalpurple}{rgb}{0.47, 0.32, 0.66}

\begin{document}

\title{The role of magnetic and rotation axis alignment in driving fast radio burst phenomenology}

\correspondingauthor{Paz Beniamini}
\email{pazb@openu.ac.il}

\author[0000-0001-7833-1043]{Paz Beniamini}
\affiliation{Department of Natural Sciences, The Open University of Israel, P.O Box 808, Ra'anana 4353701, Israel}
\affiliation{Astrophysics Research Center of the Open university (ARCO), The Open University of Israel, P.O Box 808, Ra'anana 4353701, Israel}
\affiliation{Department of Physics, The George Washington University, 725 21st Street NW, Washington, DC 20052, USA}

\author{Pawan Kumar}
\affiliation{Department of Astronomy, University of Texas at Austin, Austin, TX 78712, USA}

\begin{abstract}
We propose a scenario that can describe a broad range of FRB phenomenology, from non-repeating bursts to highly prolific repeaters. Coherent radio waves in these bursts are produced in the polar cap region of a magnetar, where magnetic field lines are open. The angle between the rotation and magnetic axes, relative to the angular size of the polar cap region, partially determines the repetition rate and polarization properties of FRBs. We discuss how many of the properties of repeating FRBs—such as their lack of periodicity, energetics, small PA swing, spectro-temporal correlation and inferred low source density are explained by this scenario. The systematic PA swing and the periodic modulation of long duration bursts from non-repeaters are also natural outcomes. We derive a lower limit of about 400 on the Lorentz factor of FRB sources applying this scenario to bursts with a linear polarization degree greater than 95\%.

\end{abstract}


\section{Introduction} 
\label{sec:intro}

Arguably, the most fundamental questions in the study of fast radio bursts (FRBs) regard the physical mechanism responsible for the emission and the underlying nature and properties of their sources.
In recent years, several important observations have provided us with tools to critically examine these issues. Variability of FRB lightcurves and the narrowness of their spectra provide important information regarding the location and therefore the origin of their radio emission, cf. \citep{BK2020,Kumar2024}. Although in some cases the fast variability of FRB light curves could result from propagation effects such as filamentation instability (e.g., \citealt{sobacchi21_modulational_instability}), it is much harder for propagation effects to convert an intrinsically broad-band radio spectrum into a narrow observed spectrum (\citealt{Kumar2024}). Therefore, the narrowness of the observed spectrum in several cases \citep{2022RAA....22l4001Z,2023ApJ...955..142Z} as well as instances of isolated and temporally very narrow pulses ($\lesssim 0.1$\,ms) provide important evidence supporting the magnetospheric origin of FRBs. An independent constraint on the location of the radio source comes from the modulation of their spectra due to scintillation in the host galaxy \citep{KBGC2024}. This technique has recently been used by \cite{Nimmo2025} to show that the emission of (apparent non-repeater) FRB 20221022A, has likely originated from within the magnetosphere of a neutron star (NS).

Another important clue to the FRB mechanism is provided by polarization observations. Many FRBs show nearly 100\% linear polarization. The recently reported S-swing of the electric field vector of the same burst mentioned above, FRB 20221022, appears to be a rare occurrence \citep{mckinven2025}. Sudden jumps in the polarization angle (PA) by almost $\pi/2$ are also found for a small fraction of bursts from the repeater FRB 20201124A \citep{Niu2024}. Although these events are rare they nevertheless offer strong support for the magnetospheric origin of FRBs and provide insights into the possible origin of the difference between repeating and non-repeating FRBs. Additional evidence for the magnetospheric origin of coherent radio emission in FRBs is provided by the highly periodic series of bursts from FRB 20191221A (\citealt{CHIMEperiodicity, Beniamini2023}), and the large total energy requirement for some prolific repeaters (as will be detailed in \S \ref{sec:energyrepeat}). The energy requirement becomes a serious concern if the ratio of X-ray to radio emission were as large as that of Galactic FRBs, as suggested by the far-away class of models for these bursts \citep{MBSM2020}. The objections raised against the magnetospheric model are primarily theoretical. These include the challenge of radio waves escaping from the magnetosphere of magnetars and the lack of a precise, self-consistent, model for coherent radio wave emission \citep{Beloborodov2021,Golbraich2023,Sobacchi2024}. However, the issue of wave escape from the magnetosphere is debated; while some papers argue that the waves cannot escape, other works show they can, particularly along open magnetic field lines under plausible physical conditions, especially when ponderomotive acceleration of plasma by a precursor burst or the head of the FRB pulse is included in the calculation \citep{Qu2022,Lyutikov2024}. Given that current observational evidences favor the magnetospheric mechanism for FRBs, we investigate some implications of this model and whether it can explain the differences between non-repeating FRBs and the class of repeaters. 

The properties of FRBs are highly diverse, with e.g. energy ranges, activity rates and burst durations that all span multiple orders of magnitude (e.g. \citealt{Zhang2023}). As a specific example, based on their volume densities, we know that only a small fraction (of order one in a million) of magnetars are highly active FRB sources \citep{LBK2022}. What properties of a magnetar are responsible for producing these powerful radio bursts and account for this wide diversity? There are only a small number of magnetar parameters that could potentially serve this role. Chief among them are perhaps the magnetic field strength, rotation rate, inclination between the rotation and magnetic axes, and the magnetar's age \footnote{Twisting of the magnetosphere by sub-surface and crustal activities likely also plays an important role (as evident by the temporal proximity of spin glitches and FRB-like bursts from the Galactic magnetar SGR 1935+2154, \citealt{Younes2023}). Some magnetars exhibit hundreds of X-ray bursts in less than an hour \citep{Younes2020}, and some highly prolific FRB repeaters (e.g. 20201124A) can produce tens of bursts within minutes to hours. The underlying central trigger for these different events might be similar and could be related to magnetospheric deformation. However, the natural timescale for the release of free energy stored in the deformed fields is the Alfv\'en crossing time, which is a few milliseconds. Describing field deformation would require introducing many additional unknown parameters and the degree to which they vary between sources and over time, which is not yet well understood and that we do not consider in this work.
Instead, we focus on exploring the implications of a component that is purely geometric and is naturally expected to vary between sources: the inclination between the magnetic and rotational axes. 
}.
As we argue in this work, both repeaters and non-repeaters typically require very strong magnetic field strengths of $B\gtrsim 10^{14}$\,G. 
At the same time, stability of the NS implies that $B\lesssim 10^{16}\mbox{ G}$ \citep{2013MNRAS.433.2445A}.
This leaves limited room for variation in $B$ to explain the broad range in FRB phenomenology. The rotational period is less strongly constrained, and for the most part only a lower limit can be placed on it. 
However, this limit is sufficiently large to argue that FRBs are almost certainly powered magnetically, and as such period variation alone cannot explain the entire gamut of FRB observations. Through this process of elimination, we arrive at the conclusion that the orientation of the magnetic axis relative to the rotation axis could be a key parameter in driving differences between repeaters and non-repeaters. These arguments, together with the discovery of large PA swings and the periodic signal from the long-duration FRB 20191221A, lead us to propose the rotating polar cap scenario for FRBs.

In \S\ref{sec:basic}, we describe the rotating polar cap scenario for FRBs. In particular, we discuss the polarization swing calculation and constrain the Lorentz factor of the source based on the observed degree of polarization.
Evidence for this model is presented in \S\ref{sec:evidence} for both non-repeating and repeating FRBs. We show that, with a very small number of underlying assumptions, the model can naturally and simultaneously account for many observed features of these bursts. We outline a few predictions of the model which can be tested by future observations in \S \ref{sec:predict} and conclude in \S\ref{sec:conclusion}.
In the appendix, we present the rationale that FRBs are powered by magnetic field dissipation rather than NS rotational kinetic energy and its implications for underlying periods and the polar cap size.

\begin{figure}
\centering
\includegraphics[scale=0.17]{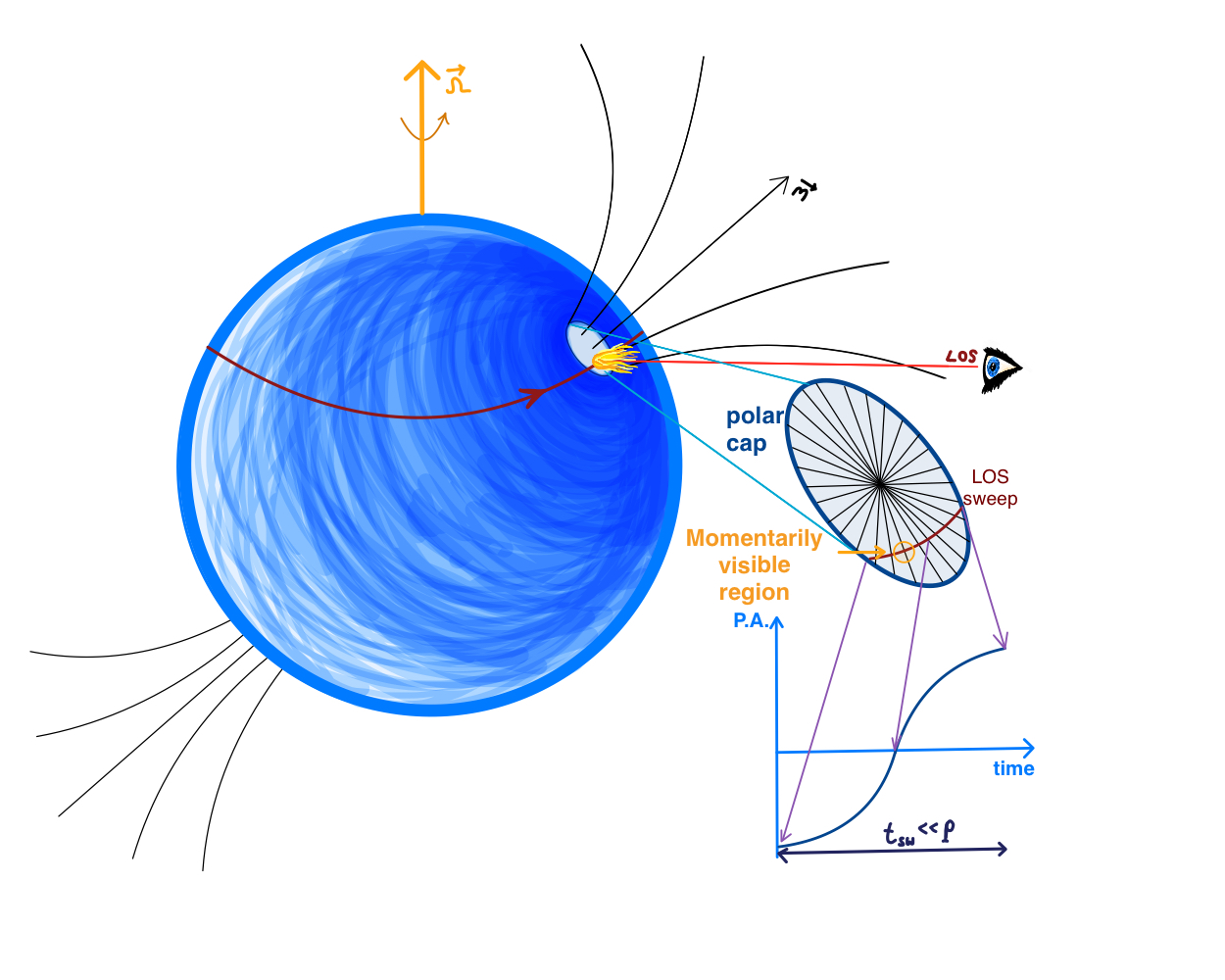}
\includegraphics[scale=0.17]{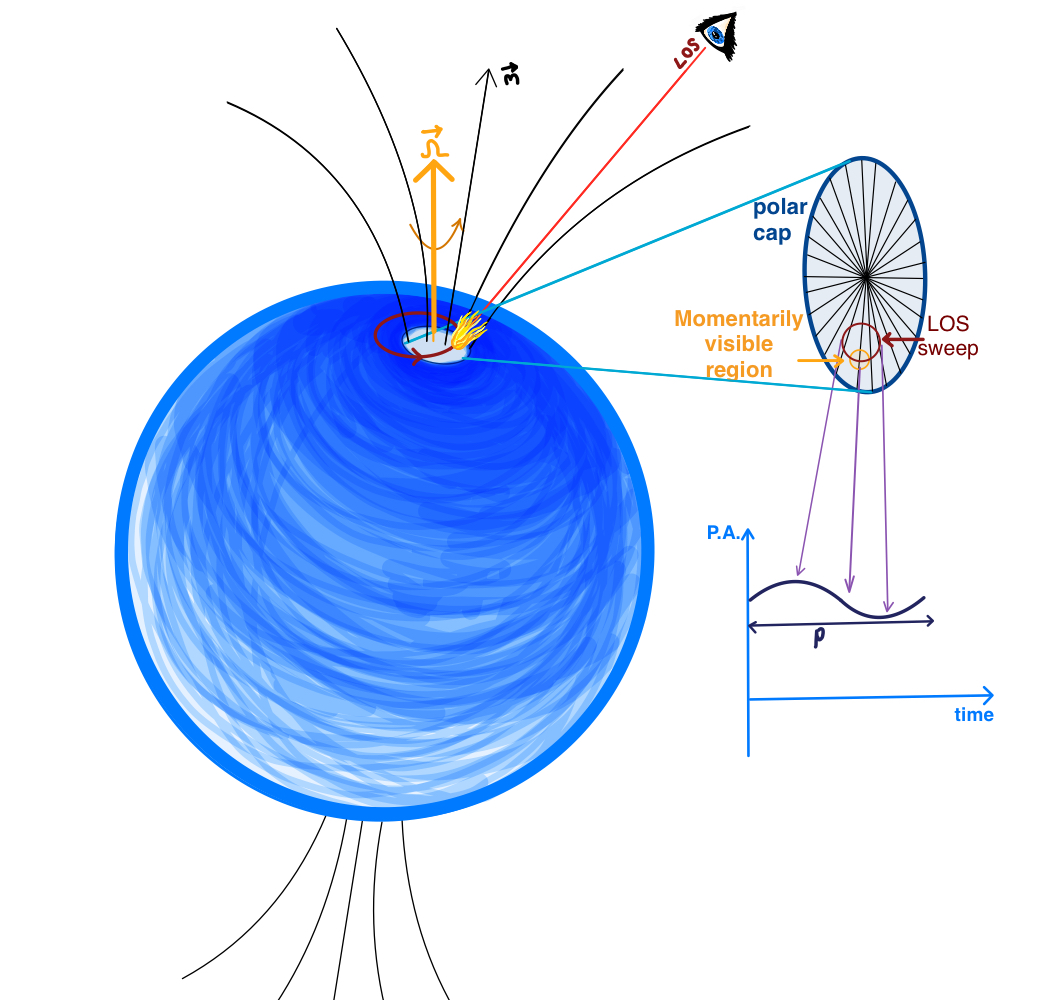}
\caption{Schematic figure representing the geometry of a rotating NS with a dipolar-like magnetic field in which the radio emission escapes towards the observer from a small patch within the polar cap. Due to relativistic beaming, the momentarily visible emitting patch has an angular size much smaller than that of the polar cap. Left: The magnetic axis, $\vec{m}$ is highly misaligned from the spin vector, $\vec{\Omega}$, i.e. the angle between them, $\alpha$, is large compared to the opening angle of the rotating beam, $\rho$. This situation can lead to large polarization angle swings, as long as the intrinsic burst duration is $\gtrsim t_{\rm sw}$, the time it takes the polar cap to sweep across the LOS. Right: Aligned rotator, i.e. $\alpha\ll\rho$. In this situation the polar cap is always in view and no underlying periodicity or strong polarization angle swings are seen.}
\label{fig:schematic}
\end{figure}

\section{Rotating polar cap scenario - basic ingredients}
\label{sec:basic}

Our model considers magnetospheric FRB emission from the open field lines near the magnetic pole (schematic figure \ref{fig:schematic}). A rotating polar cap has been extremely successful in describing pulsar phenomenology \cite{RC1969}. However, as the energy source powering the FRB emission is almost definitely magnetic (see appendix \S \ref{sec:energy}) --  as opposed to pulsars, which are powered by the NS spin kinetic energy -- this model can be thought of as a hybrid pulsar-magnetar model \cite{BK2023}.
We begin by outlining some of the basic ingredients of this model. For quantitative estimates we focus on a dipolar magnetic field geometry. This choice is done primarily for the sake of clarity and convenience. Our results will be qualitatively unchanged and quantitatively similar for other field geometries.

Consider a magnetar, with a magnetic axis, $\vec{m}$, that is misaligned from the rotation axis, $\vec{\Omega}$, by an angle $\alpha$. The former axis is rotating around the latter with the rotational period of the NS, $P$. We will consider emission from relativistic plasma at a radius $r$ from the NS center.
Since the plasma is highly relativistic (see \S \ref{sec:poldeg} for more details) and since its motion is directed along magnetic field lines, the emission seen by the observer is centered on a point in the magnetosphere that satisfies the condition $\vec{B}(\theta_0,\phi_0)\parallel \hat{n}_{\rm obs}$, where $\theta_0, \phi_0$ are the the latitude and longitude of the central emission point defined relative to a frame rotating with the NS, such that $\vec{m} \parallel \hat{z}$. Assuming, for concreteness, a dipole field geometry, we have (in the same frame) $\vec{B}(\theta,\phi)\propto (1.5\sin 2 \theta \cos \phi, 1.5 \sin 2 \theta\sin \phi, 3\cos^2\theta-1)$. With no loss of generality, we can take the observer to be in the $x-z$ plane, such that $\hat{n}_{\rm obs}=(\sin \beta, 0,\cos \beta)$ (i.e. $\beta$ is the impact parameter of the observer's LOS relative to the magnetic axis). The condition $\vec{B}(\theta_0,\phi_0)\parallel \hat{n}_{\rm obs}$ is then equivalent to
\begin{equation}
\label{eq:emcenter}
  \sin(\beta)=3\sin(2\theta_0-\beta)  \quad \& \quad \phi_0=0
\end{equation}
For $\beta\ll 1$, the result becomes simply $\theta_0=\frac{2}{3}\beta, \phi_0=0$. The result can then be transformed to a frame with $\Omega\parallel \hat{z'}$ by first rotating $\vec{r}_0=(\sin \theta_0 \cos \phi_0,\sin \theta_0 \sin \phi_0,\cos \theta_0)$ by an angle $\alpha$ relative to $\hat{y}$ and then rotating by $\Omega t$ relative to $\hat{z}'$, leading to $\vec{r_0'}\propto (\sin(\theta_{\rm 0}+\alpha)\cos \Omega t, \sin(\theta_{\rm 0}+\alpha)\sin \Omega t,\cos (\theta_{\rm 0}+\alpha))$.

The polar cap radius at distance $r$ from the NS is $r_{\rm c}=r(2\pi r/cP)^{1/2}$. Therefore, the half-opening angle of the polar cap region is
\begin{equation}
\label{eq:thetapc}
    \theta_{\rm pc}=\frac{r_{\rm c}}{r}=\left(\frac{r}{R_{\rm LC}}\right)^{1/2}= {0.014\, \rm rad}\,  r_6^{1/2}P_0^{-1/2}
\end{equation}
where $R_{\rm LC}$ is the light cylinder radius and throughout the paper we use the convention $Q_x=Q/10^x$ in cgs units.
We define the `sweep time', as the time it takes the polar cap to sweep across the observer LOS. It is given by 
\begin{equation}
\label{eq:tsw}
   t_{\rm sw}\!=\!\frac{WP}{2\pi}\!\approx \!\frac{P \sqrt{\rho^2-\beta^2}}{\pi \sqrt{\sin \alpha \sin(\alpha+\beta)}}\!\approx \!{7\,\rm ms}\,  r_6^{1/2} P_0^{1/2}g(\alpha,\beta,\frac{\beta}{\theta_{\rm pc}})
\end{equation}
where $\rho$ is the half-opening angle of the emitting beam in the observer frame (as shown above $\rho\approx (3/2)\theta_{\rm pc}$ for $\theta_{\rm pc}\ll 1$) and where $W$ is the angular width of the segment of the polar cap intersecting with the LOS. In deriving this expression we have assumed that $\beta\leq \rho$ is required for the observer to be able to see emission from the polar cap.
An exact expression for $W$ is given by $\sin^2(W/4)=(\sin^2(\rho/2)-\sin^2(\beta/2))/(\sin \alpha \sin(\alpha+\beta))$ \citep{Gil1984,LK2012}. A more intuitive, approximate equation (valid in the small angle limit) is given by $W\approx 2\sqrt{\rho^2-\beta^2}/\sqrt{\sin \alpha \sin(\alpha+\beta)}$. Finally $g(\alpha,\beta,\beta/\theta_{\rm pc})\equiv \sqrt{1-\beta^2/\rho^2}/\sqrt{\sin \alpha \sin(\alpha+\beta)}$ is a dimensionless function, which (except for geometrically fine tuned scenarios) is  of order unity.

We denote the intrinsic duration of a burst as $t_{\rm br}$. The observed duration will depend on the ordering of this timescale relative to $t_{\rm sw}$ and $P$ (as well as on $\alpha$ relative to $\theta_{\rm pc}$). The situation is summarized in table \ref{tab:summary}.
In particular, we note that when $\alpha$ is small, the burst duration is always determined by $t_{\rm br}$. Alternatively, when $\beta\approx \theta_{\rm pc}$ is small, so is $t_{\rm sw}$, and therefore $t_{\rm br}$ would be larger than the sweep of the magnetic axis.

\begin{table}
\tiny
\setlength{\tabcolsep}{5pt}
\begin{center}
\begin{tabular}{lccccc}
 & spike duration &  FRB duration & polarization degree &  polarization swing & comments \\
\hline \hline
$\alpha\gtrsim \rho$ & &  & & & $f_{\rm b,eff}=\rho\sin \alpha $\\
\hline 
$t_{\rm br}<t_{\rm sw}<P$ & $t_{\rm br}$ & $t_{\rm br}$ & $\sim 1$ & $\ll 1$ &Single spike  \\
$t_{\rm sw}<t_{\rm br}<P$ & $t_{\rm sw}$ & $t_{\rm sw}$ &  $\sim 1$ if $|\beta|\sim \rho$ or $t_{\rm res}\ll t_{\rm sw}$ & $\sim 1$ if $|\beta|\lesssim \rho/2$ and $t_{\rm res}\ll t_{\rm sw}$ & Single spike\\
 &  &  & $\ll1 $ else  & $\ll 1$ else& \\
$t_{\rm sw}<P<t_{\rm br}$ & $t_{\rm sw}$ & $t_{\rm br}$ &  $\sim 1$ if $|\beta|\sim \rho$ or $t_{\rm res}\ll t_{\rm sw}$ & $\sim 1$ if $|\beta|\lesssim  \rho/2$ and $t_{\rm res}\ll t_{\rm sw}$ &  Periodic spikes \\
 &  &  & $\ll1 $ else  & $\ll 1$ else & (pol. results are per spike)\\
\hline
$\alpha\ll \rho$ & &  & & & $f_{\rm b,eff}=\frac{1}{2}\rho^2$\\
\hline
any & $t_{\rm br}$ & $t_{\rm br}$ & $\sim 1$ if $|\beta|\gtrsim 2\alpha$ or $t_{\rm res}\ll P/2$ & $\ll 1$ if $|\beta|\gtrsim 2\alpha$ or $t_{\rm res}\gtrsim P/2$ & Single spike \\
 &  &  & $\ll 1$ else  & $\sim 1$ else & \\
\hline
\end{tabular}
\end{center}
\caption{Relation between observed properties (duration, polarization, effective beaming) as compared with the intrinsic ones, for the different geometries and different orderings of the characteristic times. The polarization results quoted in the table are valid when the emitting particles' Lorentz factor is sufficiently large (see Eq. \ref{eq:Gammalimpol}).} 
\label{tab:summary}
\end{table}

\subsection{Rotating polar cap scenario: polarization fraction and constraint on source Lorentz factor}
\label{sec:poldeg}

Many FRBs show a large degree of polarization, $1-\Pi\ll 1$. At first glance, this might appear surprising in the context of the rotating beam model, considering that the PA direction is typically assumed to be along (or at a fixed angle relative to) the curvature vector of the local magnetic field ($\vec{B}$) projected onto the plane of the sky. This PA changes significantly across the polar cap, as shown explicitly in \S\ref{sec:PAswing}, thereby reducing the value of $\Pi$ if emission from different points in the polar cap is uncorrelated and a large portion of the polar cap is emitting and visible to the observer at a fixed time.

Therefore, for $\Pi \approx 1$ to occur, one of two conditions must be satisfied: either the entire polar cap emits coherently, allowing the electric field vectors from different points to combine and produce the observed flux and high polarization degree, or the visible portion of the polar cap at a fixed observer time is much smaller than $\theta_{\rm pc}$ due to relativistic beaming, where a large fraction of the polar cap is emitting but only a small region around the central emission point, $\theta_0, \phi_0$ (\S\ref{sec:basic}), is beamed toward the observer\footnote{A large $\Pi$ can alternatively be explained if the angular size of the region producing bursts, $\theta_{\rm br}$ is much smaller than $\theta_{\rm pc}$. 
This situation can also lead to a large polarization angle swing, independently of the Lorentz factor of the emitting material. However, in this case, if $\Gamma<\theta_{\rm br}^{-1}$ then there is no beaming correction applied to the observed fluence and the inferred energy from each observed burst is the true emitted energy, $f_b\to 1$. This can quickly lead to an energy crisis (see \S \ref{sec:energyrepeat} for details). For this reason we focus in this work on the case for which $\Gamma>\theta_{\rm br}\approx \theta_{\rm pc}$.}. To understand the degree of beaming in this situation, we begin with the simplest example, which is a non-rotating NS with a dipole field geometry. The corrections due to rotation are discussed below and can be shown to typically be small for the values of $\Pi$ observed in FRBs and the values of $P$ inferred for the rotating polar cap scenario. Furthermore, the assumption of dipolar field geometry is conservative. This is because any higher order multipolar components, will lead to a decrease in the angular extent of the region over which the PA changes significantly. This in turn will require an even larger Lorentz factor in order to reproduce the high degree of observed polarization.

The entire polar cap is unlikely to radiate coherently if the source is moving relativistically with a large Lorentz factor, $\Gamma$, as suggested by the high brightness temperature of FRBs, especially if the angular size of the polar cap is larger than $1/\Gamma$, as required by the PA-swing data. From now on, we assume that the portion of the polar cap visible to the observer at a fixed time consists of multiple patches, each of which radiates coherently, but the radiation from different patches adds incoherently.

As in \cite{RC1969}, we consider the polarization direction from a point $(\theta,\phi)$ to be along the magnetic axis projected in the plane of the sky or the plane perpendicular to the radial vector terminating at the point $(\theta,\phi)$\footnote{The results are unchanged if the PA is rotated within this plane by any fixed amount relative to the projected magnetic axis, such as e.g., when the polarization is perpendicular to the magnetic axis projected on the sky.}.
The direction of the electric field for radiation from point ${\hat r}$ is along $\hat{m}\times \hat{r} = \sin\theta( -\sin\phi\, \hat x + \cos\phi \, \hat y)$, if we take the magnetic axis, $\hat m$, to be along $\hat z$.  Thus, the contribution of the radiation to Stokes parameter Q and U are proportional to $\cos2\phi$ \& $\sin2\phi$ respectively.
The Stokes parameters transform from the local to the observed frame according to $I\to \mathcal{D}^2I, (Q^2+U^2)^{1/2}\to \mathcal{D}^2(Q^2+U^2)^{1/2},V\to \mathcal{D}^2V$ \citep{Cocke1972} where 

\begin{equation}
    \mathcal{D}=[\Gamma(1-\beta \cos \theta_{\rm B})]^{-1} 
\end{equation}
is the Doppler factor, and $\theta_{\rm B}$
is the angle between the emitters' velocity vector and the observer LOS, which for the case of plasma moving along magnetic field lines is given by $\cos\theta_{\rm B} = \hat{n}_{\rm obs}\cdot\hat{B}(\theta,\phi)$. 
Assuming the polarization is time resolved (which is the best case scenario for purposes of maximizing the polarization degree, see below), we can take, as above, the observer to be in the x-z plane.
Under our assumptions, $V=0$ and for the choice of axes described above, $U=0$ as well because $\sin(2\phi)$, averaged over $\phi$, is zero. Therefore, the general expression for the polarization degree, when averaged over different $\theta$ and $\phi$, is given by:
\begin{equation}
\label{eq:Pigeneral}
    \Pi\equiv \frac{Q}{I}=\frac{\int  \mathcal{D}^2 \cos 2\phi \sin \theta d\theta d\phi}{\int  \mathcal{D}^2 \sin \theta d\theta d\phi}.
\end{equation}
where, for the sake of concreteness, we have assumed a constant emissivity per unit solid angle across the emitting surface.
We define the polar angular difference relative to the emission center point, $\Delta \theta=\theta-\theta_0$. We also note that since for the rotating polar cap scenario $\theta_0\lesssim \theta_{\rm pc}$, it is safe to assume (see Eq. \ref{eq:thetapc}) that $\theta_0\ll 1$ and therefore that $|\Delta \theta| \ll 1$ and (as shown below and verified by full numerical calculation, see \S \ref{sec:beamingandpolnumerical}), $|\phi|\ll 1$. In this limit, the expression in Eq. \ref{eq:Pigeneral} can be simplified significantly.
Since the Doppler factor decreases rapidly when $\theta_B > 1/\Gamma$ (and since $\Gamma\gg 1$), most of the contribution to the integrals comes from a small solid angle, and the result is that $\Pi \approx \cos 2\phi_\Gamma$, where $\phi_\Gamma$ is the value of $\phi$ that corresponds to $\theta_B\sim \Gamma^{-1}$. Expanding $\cos\theta_B$ in small parameters,
\begin{equation}
\label{eq:SeriescosthetaB}
    \cos\theta_B \equiv \hat{n}_{\rm obs}\cdot\hat{B}(\theta,\phi) \approx 1 - {\theta_B^2\over 2} \approx 1-\frac{9}{8}\Delta \theta^2 -\frac{9}{8} \phi^2\theta_0^2
\end{equation}
The Doppler factor can also be expanded as $\mathcal{D} \approx 2\Gamma/[1 + (\Gamma\theta_B)^2]$. Combining the expressions for $\mathcal{D}$ and $\theta_B$, and using the fact that most of the contribution to the integral in equation (\ref{eq:Pigeneral}) come from where $\mathcal{D}\gtrsim \Gamma$, we find that $\phi_\Gamma^2 \approx 1/[2\theta_0^2\Gamma^2]$, and therefore
\begin{equation}
    \epsilon \equiv 1 - \Pi \approx 1- \cos2\phi_\Gamma \approx {1\over \theta_0^2 \Gamma^2}.
\end{equation}
The approximation in the derivation of this equation breaks down when $\theta_0=0$. 

Recalling that $\theta_0\lesssim \theta_{\rm pc}$, one can place a lower limit on $\Gamma$ for a given observed value of $\epsilon$, which is\footnote{The factor of $1.7$ comes from doing the full numerical integration.}
\begin{equation}
\label{eq:PoldegGamma}
    \Gamma\gtrsim (\epsilon/1.7)^{-1/2} \theta_{\rm pc}^{-1}\sim 930\, \epsilon_{-2}^{-1/2} r_6^{-1/2} P_0^{1/2}
\end{equation}
Note that this is a strict lower limit, as any other effect not included in this simple estimate will only tend to reduce the observed polarization further. In particular, this is the case, when considering the effect of pulsar rotation as we briefly outline below.

To account for the degradation of the polarization degree due to rotation, we consider first a source that is 100\% linearly polarized and, whose PA is changing over time, PA=PA$(t)$. The typical timescale over which the PA changes significantly is of order $t_{\rm sw}/\mbox{PAS}_{\rm pc}$ (where PAS$_{\rm pc}(\alpha, \beta)$ is defined as the overall change in PA, or the polarization angle swing (PAS), as the polar cap comes in and out of view, see \S \ref{sec:PAswing}). Depolarization will happen due to finite temporal resolution of the detector, $t_{\rm res}$, i.e. during a single epoch of observations the PA would have rotated such that the averaged $\Pi$ becomes less than unity. In the limit $t_{\rm res} \!\ll \! t_{\rm sw}/\mbox{PAS}_{\rm pc}$ the polarization degree is $\Pi=1-\frac{1}{6} \mbox{PAS}_{\rm pc}^2(t_{\rm res})\approx 1-\frac{1}{6} \left(\frac{ \mbox{PAS}_{\rm pc}\cdot t_{\rm res}}{t_{\rm sw}}\right)^2$ \citep{BKN2022}.

Combining the effects of rotation and beaming, we can express the deviation of the polarization degree from unity as
\begin{equation}
\label{eq:Gammalimpol}
    \epsilon=\max\left(\frac{1}{2}(\Gamma \theta_{\rm pc})^2,\frac{1}{6} \left(\frac{ \mbox{PAS}_{\rm pc} \cdot t_{\rm res}}{t_{\rm sw}}\right)^2\right)
\end{equation}
i.e., the two effects can be treated separately, and the larger of the two would dominate the  depolarization. How significant is the depolarization due to rotation? Taking as an example FRB 20221022A, the polarization analysis by \cite{mckinven2025} is done using a temporal resolution $t_{\rm res}=20\mu \mbox{s}$. As discussed in \S \ref{sec:evidencePC}, in the context of the rotating polar cap scenario, we estimate that $t_{\rm sw}\approx 5$\,ms for this burst. Thus the degradation of polarization due to rotation is at most $\epsilon_{\rm rot}\lesssim 3\times 10^{-6}$. This should be contrasted with the observed $\epsilon=0.05\pm0.03$ for the same burst, which is clearly orders of magnitude larger. We conclude that rotation is in practice not constraining for the degree of polarization. The high degree of observed polarization does, however, provide very useful limits on the emitting particles' Lorentz factor as shown by Eqns. \ref{eq:PoldegGamma} and \ref{eq:Gammalimpol}.

\subsection{Polarization swing}
\label{sec:PAswing}
\begin{figure}
\centering
\includegraphics[scale=0.25]{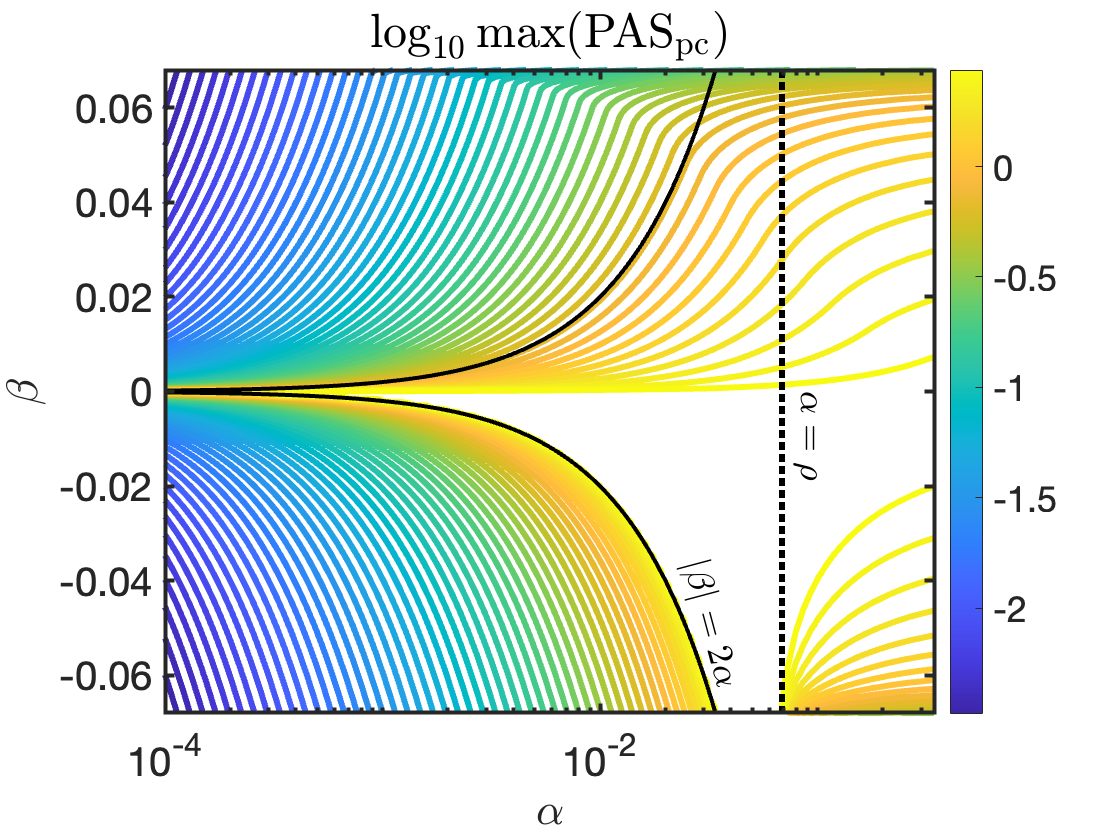}
\caption{The maximum polarization angle swing incurred as the observer line of sight crosses the polar cap (PAS$_{\rm pc}$). For $\alpha\gtrsim \rho$ (misaligned rotators), PAS$_{\rm pc}$ is of order unity unless $\beta\approx \rho$. For $\alpha\lesssim \rho$ (aligned rotators), PAS$_{\rm pc}$ is typically $\ll 1$, but order unity PAS$_{\rm pc}$ is possible when $|\beta|\lesssim 2\alpha$.}
\label{fig:PASpc}
\end{figure}

As discussed in \S \ref{sec:poldeg}, the PA is given by the projection of the magnetic axis in the sky. Consider a Cartesian coordinate system with its z-axis along the rotation axis. The co-latitude of the magnetic axis in this coordinate system is $\alpha$, and the observer is at $(\alpha+\beta$). The angular speed of the NS is $\Omega$. The time-dependent magnetic axis is: $\hat m(t) = (\sin\alpha \cos\Omega t, \sin\alpha \sin\Omega t, \cos\alpha$). The observer coordinates are $\hat n_{\rm obs} = [\sin(\alpha+\beta), 0, \cos(\alpha+\beta)]$. The direction of the radiation electric field is, therefore, 
\begin{equation}
    \vec E = \hat m(t) \times \hat n_{\rm obs} = \hat x_\perp \sin\alpha \sin\Omega t + \hat y \left[ \cos\alpha \sin(\alpha+\beta) - \sin\alpha \cos(\alpha+\beta) \cos\Omega t\right] \equiv E_{x_\perp} \hat x_\perp + E_y \hat y
\end{equation}
where $\hat x_\perp \equiv \cos (\alpha+\beta) \hat x - \sin(\alpha+\beta)\hat z$, is a vector perpendicular to the line of sight; $\hat x_\perp$ \& $\hat y$ are orthogonal vectors, perpendicular to $\hat n_{\rm obs}$ in the plane of the sky. Thus, the PA at time $t$, wrt $\hat y$, is given by
e.g. \cite{RC1969, LM1988},
\begin{equation}
\label{eq:PAS}
    \tan (\mbox{PA})= {E_{x_\perp}\over E_y } =\frac{\sin \alpha \sin (\Omega t)}{\sin (\alpha+\beta)\cos \alpha-\cos(\alpha+\beta)\sin \alpha \cos(\Omega t)}
\end{equation}
where with no loss of generality we have chosen $t=0$ such that $\vec{\Omega}, \vec{m}, \hat{n}_{\rm obs}$ are all in the same plane. Using Eq. \ref{eq:PAS} we can calculate the PAS across the polar cap, $\mbox{PAS}_{\rm pc}(\alpha,\beta)$. The results are shown in Fig. \ref{fig:PASpc}. and can be intuitively understood in limiting cases as discussed below.

\begin{figure}
\centering
\includegraphics[scale=0.5]{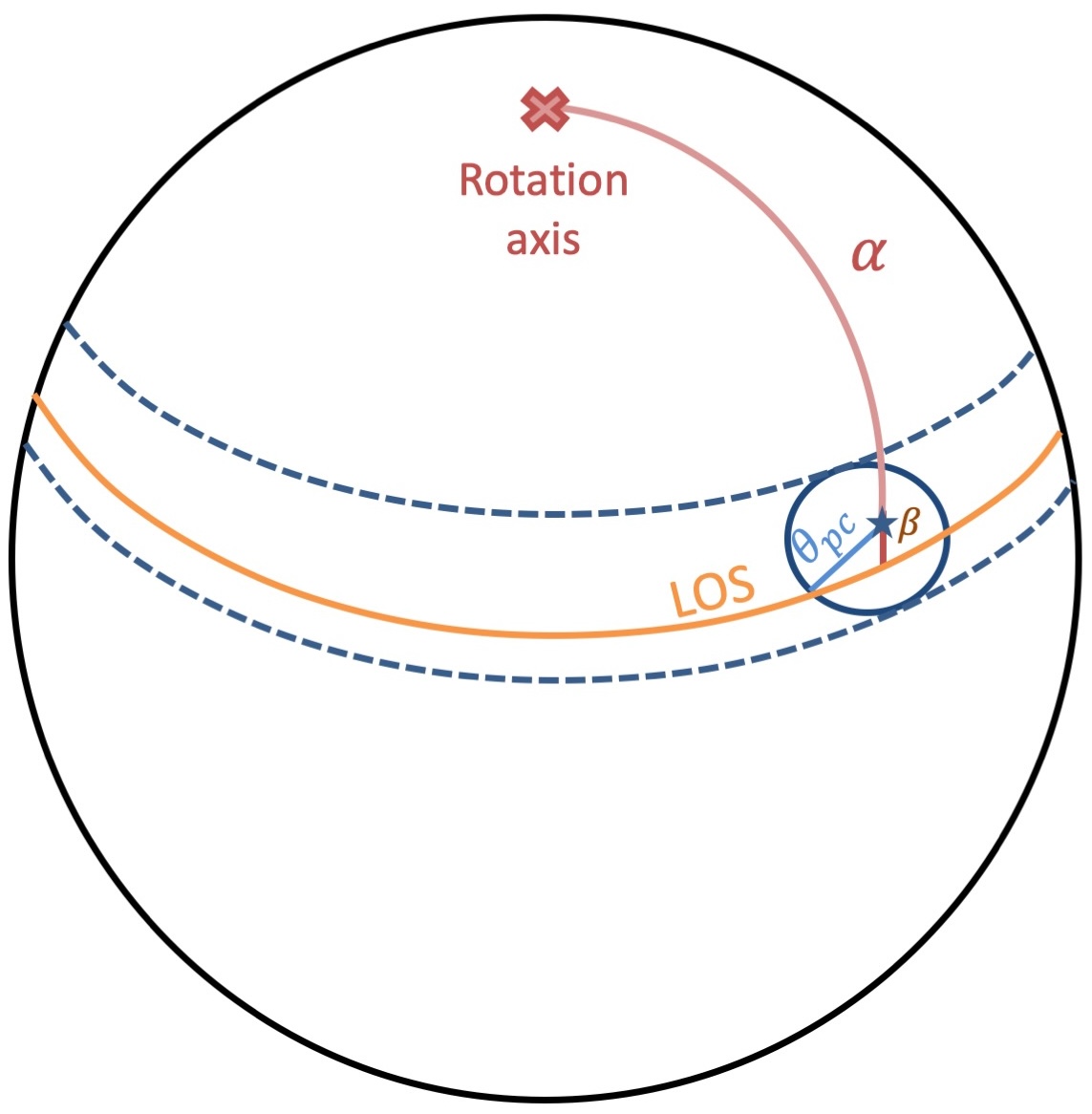}
\includegraphics[scale=0.5]{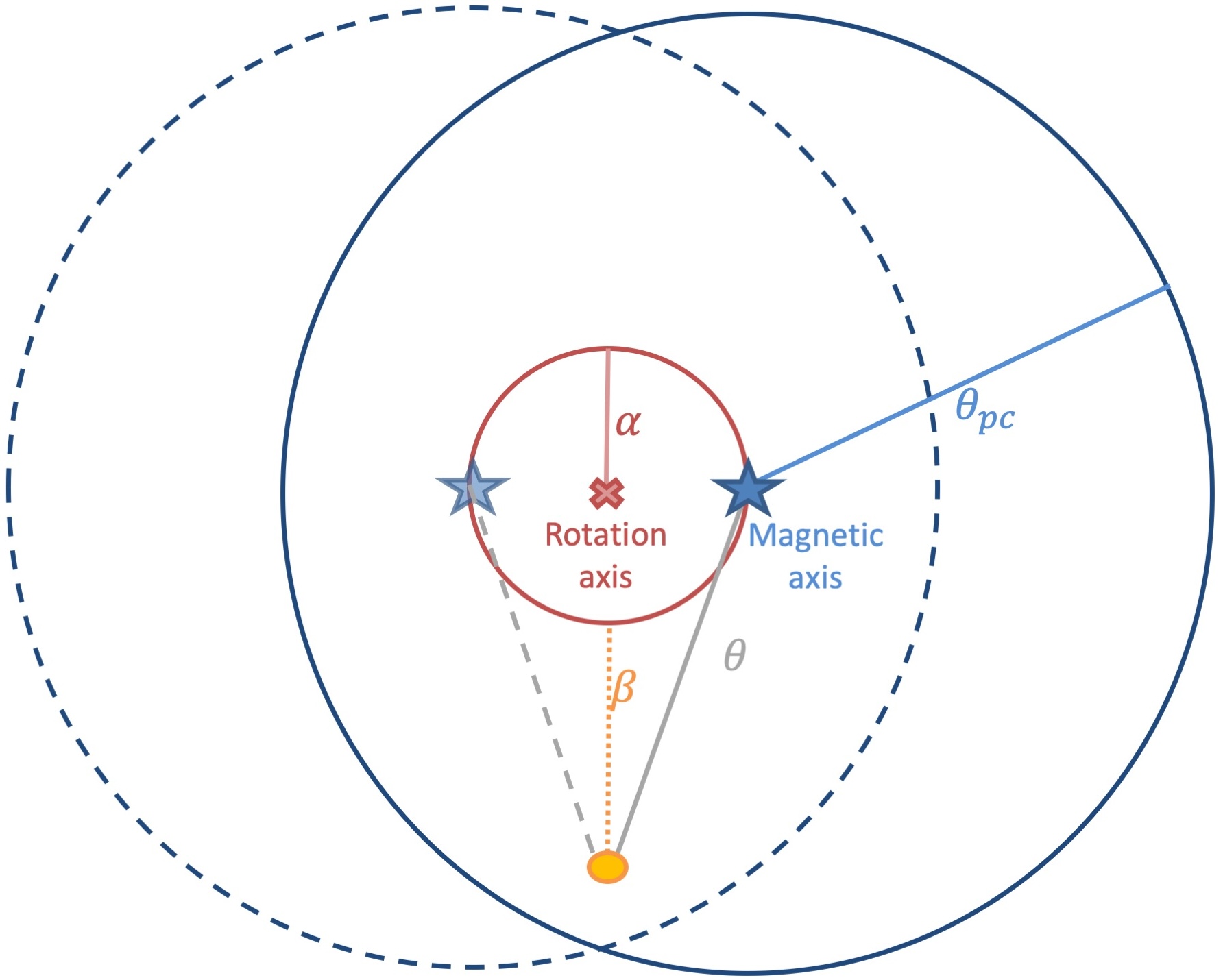}
\caption{Top: schematic figure representing the case $\alpha>\theta_{\rm pc}$, in which the magnetic axis (represented by a star) misalignment from the rotation axis (represented by a cross) is large compared to the polar cap size. The observer LOS is represented by an orange line and has an angular impact parameter of $\beta$ relative to the magnetic axis. Bottom: Face on view of the geometry for the case $\alpha<\theta_{\rm pc}$. In this case the observer is at an angular offset $\beta$ relative to the rotation axis (cross), and is always within the polar cap region. The polar cap region is shown for two points in time, removed by half an orbital period.}
\label{fig:schematic2}
\end{figure}

When $\alpha\gtrsim \rho$, the polar-cap rotates in and out of the line of sight (see top panel of Fig. \ref{fig:schematic2}). The overall PAS$_{\rm pc}$ is then related to the impact parameter $\beta$. Taking $\alpha\approx \pi/2, \beta\ll 1$, $\tan\mbox{PAS}_{\rm pc}\approx \frac{\rho}{\beta}\sqrt{1-\frac{\beta^2}{\rho^2}}$. In particular, $\tan(\mbox{PAS}_{\rm pc})\ll 1$ for $\beta\approx \rho$, it increases to order unity for $\beta\approx \rho/2$ and becomes $\gg 1$ for smaller values of $\beta$. If the burst duration is shorter than the polar-cap sweep time, the observed PAS depends also on the ratio of those two. The observed PAS can then be approximated as $\mbox{PAS}\approx \max(\frac{\rho}{\beta}\sqrt{1-\frac{\beta^2}{\rho^2}},1) \min(t_{\rm br}/t_{\rm sw},1)$. In particular we see that for typical values of $\beta\approx \rho/2$, the PAS is large (i.e., of order one radian).

Alternatively, if $\alpha\ll \rho$ (see bottom panel of Fig. \ref{fig:schematic2}), the LOS is typically always within the polar-cap. In this situation, we denote by $\theta\approx \sqrt{\alpha^2+(\alpha+\beta)^2}$ the angular difference between the LOS and the magnetic axis, at a given instance. This allows us to estimate the PAS as $\mbox{PAS}\approx \min[(2\alpha/\sqrt{\alpha^2+(\alpha+\beta)^2}),1] \cdot \min[2t_{\rm br}/P,1]$. In other words the PAS is small unless the observer LOS is offset from the rotation axis by a degree of order $|\beta|\lesssim 2\alpha\ll \rho$ {\it and} the FRB duration is of order half the NS period (or more).
As a result, we expect most repeater bursts to exhibit a small PAS. Nonetheless, large PAS are still possible and could be very informative regarding the required underlying conditions.

\section{ Evidence from FRBs for a rotating polar cap scenario}
\label{sec:evidence}

Multiple independent lines of evidence bolster the rotating polar cap scenario and suggest that it might naturally explain a wide plethora of FRB observations for non-repeating and repeating FRBs. In particular we associate non-repeating FRBs with misaligned rotators ($\alpha\gg \rho$) and highly prolific repeater FRB sources with aligned rotators ($\alpha \lesssim \rho$).
We summarize below the observational support for the rotating polar cap scenario focusing on those two extreme ends of the $\alpha/\rho$ distribution. See \S \ref{sec:statrepetitionrates} for a discussion of intermediate cases.

\subsection{Non-repeating FRBs as misaligned rotators}
\label{sec:evidencePC}

The data for a few non-repeating FRBs are best explained by the model we have proposed here. The support the data for these bursts are discussed individually. Despite the underlying arguments being independent, the physical requirements between all these FRBs are remarkably similar, thus leading credence to the underlying model.

\begin{itemize}
\item {\bf Periodic burst} - FRB 20191221A. The long duration ($t_{\rm FRB}\sim 3$\,s) non-repeating, FRB 20191221A exhibits a very precisely periodic radio  signal with $P=216.8\pm 0.1$ ms \citep{subsecPCHIME}. Individual pulses for this burst were found to be narrow with a duty cycle of about $\eta=0.018$. As we have argued in \cite{BK2023}, this periodic modulation of the flux can naturally be explained by emission from the rotating polar cap of a highly magnetized NS. In particular, we have shown that: (i) the high accuracy of the periodicity ($\Delta P/P<10^{-3}$), (ii) the small duty cycle, $\eta\ll 1$ and (iii) the lack of higher harmonics, rule out alternative origins for the periodicity, such as crustal oscillations. Within the context of a rotating beam model, the duty cycle is given by $\eta=W/(2\pi)=t_{\rm sw}/P$ (see table \ref{tab:summary}). 
Eq. \ref{eq:tsw} shows that for the observed value of $P=0.22$\,s and for an order unity value of the combination $r_6^{1/2} g$ (see \S \ref{sec:basic}), the observed duty cycle of $\eta=0.018$ is naturally reproduced.

\item {\bf PA swing} - FRB 20221022A shows nearly 100\% linear polarization, and the PA varies with time, and has the familiar S-swing of the PA that pulsars show as reported by \cite{mckinven2025}.  This PA variation with time is naturally explained if a bright radio patch rotates into and then out of the line of sight (see Fig. \ref{fig:schematic}). Such PASs are commonly observed in radio pulsars, whose emission is thought to originate from the polar cap of a NS with a magnetic axis that is misaligned relative to the rotation axis \citep{RC1969}. If we associate the duration of FRB 20221022A $t_{\rm FRB}=2.5$\,ms with $t_{\rm sw}$, we find that the spin period of the NS is
\begin{equation}
\label{eq:P221022}
    P\approx 0.13 \left(\frac{t_{\rm sw}}{2.5\mbox{ ms}}\right)^2g^{-2} r_6^{-1}\mbox{ s}.
\end{equation}
We note that the orthogonal jump in the PA reported for FRB 20201124A is due to a distinct mechanism, and that is discussed in \citep{QuZhang2023}.

\item {\bf Twin FRBs} - \cite{Bera2024} reported the discovery of FRB 20210912A  which appears to be a `twin' of the previously reported FRB 20181112A \citep{Prochaska+19,Cho2020}. Both (apparent non-repeater) FRBs show two short ($\sim 0.1-0.3$\,ms) spikes separated by a longer gap ($\sim 0.8-1$\,ms). The normalized lightcurves match closely with each other. Furthermore, both bursts show a change of the PA by $\sim 30^{\circ}$ within a spike. For FRB 20210912A, the quality of the data is good enough to observe a sweep-like evolution of the PA during the first spike, which is then roughly inverted in the second spike. The authors interpret the PA swing in the context of the pulsar rotating vector model. However, they associate the gap duration with the spin period. Such a short spin period is highly unlikely, considering that a NS with such a spin will spindown rapidly on a time of a few minutes for a magnetar strength magnetic field. Moreover, a magnetar with spin as short as this would be  embedded inside a dense nebula, due to a magnetar wind, which most likely would be opaque to the FRB radio waves \citep{KBGC2024}. However, these observations are nicely accounted for by the picture presented above by associating the gap duration with $t_{\rm sw}$. This naturally accounts for the timescale with no need for a physically-challenged extremely short period, and indeed matches the values inferred for the FRBs mentioned above. Furthermore, a `hollow cone' type emission (as often considered in the pulsar literature, see \citealt{RS1975}) can naturally account for the fact that the two spikes show a `mirror image' evolution of the PA temporal evolution.

\end{itemize}

\subsection{Prolific repeaters as aligned rotators}
\label{sec:repeatersaligned}

How do extremely prolific FRB repeaters fall into the rotating polar cap picture described above?
We propose that highly active repeaters can originate from aligned rotators - magnetars for which $\alpha\lesssim \rho$, i.e. the rotation axis lies close to or within the magnetic polar cap of the NS.
This proposal can readily explain several otherwise puzzling properties of prolific FRB repeaters, as described below.
We discuss two variants of this possibility and divide the discussion accordingly below.

\subsubsection{Young magnetars with $\alpha\ll \rho$}
\label{sec:Youngmags}
Motivated by the high activity rate of prolific repeaters as well as the large rotation measures \citep{Hilmarsson2021a,Anna-Thomas2023} and the persistent emission \citep{Chatterjee+17,Niu2022,Bruni2023} that are sometimes observed in association with them, it is natural to expect that prolific repeaters could represent the young end of magnetars' age range. As discussed in \S \ref{sec:energy} the age of R1 is likely to be at least 30 years, and is probably $\gtrsim 100$\, yrs considering the constraints on the evolution of its DM. This implies (Eq. \ref{eq:Plim}) a spin period of $P\gtrsim 0.1\mbox{ s}$, we therefore use the latter value as typical for this scenario.
\begin{enumerate}
\item {\bf Underlying periodicity.} Perhaps the most notable feature of aligned rotators, is that the polar cap always overlaps with the LOS towards the observer. As a result, the arrival time intervals will be governed by the mechanism producing the bursts and will not reflect the underlying periodicity of the NS. Indeed, as mentioned in \S \ref{sec:energy} this is consistent with burst arrival time statistics from prolific repeaters \citep{2018ApJ...866..149Z,Cruces2021,2021Natur.598..267L,2023ApJ...955..142Z,2024SciBu..69.1020Z,Sheikh2024}. However, in this scenario, the long time scale periodicity on scales of tens or hundreds of days seen in R1 and R3 \citep{CHIMEperiodicity,Pastor-Marazuela2020,Rajwade2020,2021MNRAS.500..448C}, should be unrelated to the rotational period and may originate from precession or an orbital period. An alternative possibility will be presented in \S \ref{sec:ULPM}.

\item {\bf Energetics.} As discussed in \S \ref{sec:energyrepeat} for $\alpha\ll \rho$ the effective beaming correction, $f_{\rm b,eff}$, is reduced by $\sim f_{\rm sw}(\alpha\approx 1)\approx \rho$ as compared to the $\alpha  \gtrsim \rho$ case. For a given set of observed bursts from a repeater source, the required energy at the source is reduced by the same factor, of order $0.07(r_6/P_{-1})^{1/2}$ (and by a factor of $\sim f_{\rm sw}(\alpha\approx 1)^2/2\approx 2\times 10^{-3}r_6/P_{-1}$ compared to the case of random bursts on the NS surface).
\item {\bf PA swing.} As shown in \S \ref{sec:PAswing}, for $\alpha\ll \rho$, the maximum PAS (during a burst and, as long as the intrinsic PA remains fixed in the plane perpendicular to $\vec{m}$, also between bursts) is typically low. This is consistent with observations of prolific repeaters \citep{Michilli+18,Nimmo2021,Pastor-Marazuela2020,Hilmarsson2021}, which find little to no PAS within repeater bursts and often relatively small change in PAS between bursts\footnote{A notable exception is FRB 20180301A, which exhibited burst-to-burst variation of the PA \citep{LuoPol}. Its bursts, however, also have a much lower polarization degree, suggesting that the observed PA variation may be dominated by propagation effects.}. We note however that within the rotating polar cap scenario, it is possible (but rare) to have large PAS from aligned sources, assuming that both the observer LOS is $|\beta|\lesssim 2\alpha \ll \rho$ and the burst duration (or interval between bursts if considering swing between bursts) is $t_{\rm br}\gtrsim P/2$.

\item {\bf Spectro-temporal properties.} By analyzing bursts from the first CHIME catalogue, \cite{Pleunis2021b} have demonstrated that bursts from repeaters are, on average, spectrally narrower and temporally broader than bursts from apparent non-repeaters. \cite{Metzger2022} have constructed a phenomenological toy model for studying the time-frequency structure of FRBs. These authors have shown that this dichotomy between repeaters and non-repeaters can naturally be explained by varying a single parameter between the two sources, $\bar{\beta}$ - the power-law index of the frequency drift rate, $\nu_{\rm c}(t)\propto t^{-\bar{\beta}}$. Namely, when $\bar{\beta}$ is small, the observed spectrum and duration are changed by a small amount as compared to the intrinsic ones, leading to bursts that are spectrally narrow and temporally broad. Instead, when $\bar{\beta}$ is large, the central emission frequency scans across the observed band, leading to a wider observed spectrum and to a duration shorter than the intrinsic one (and potentially shorter also than $t_{\rm sw}$).
Indeed, this interpretation fits naturally in the context of the geometrical framework suggested in the present work. For significantly misaligned non-repeater sources, the LOS scans across a large segment of the projected NS surface, with widely varying magnetic field orientations (and potentially a range of emission heights). Instead, for aligned active repeaters, the LOS intersection is almost fixed relative to the NS, and therefore so is the observed emission point (see Fig. \ref{fig:schematic2}). Since it is expected that the central frequency of emission evolves strongly with, e.g., the magnetic field strength, it likely has a strong dependence on the observable emission point. This then naturally leads to effectively large $\bar{\beta}$ for non-repeaters and vice versa, as required to explain the observed data.

\item {\bf Source densities.} The volumetric density of highly active sources such as R3, is very low $7\lesssim n_{\rm obs}(R3)\lesssim 430 \mbox{ Gpc}^{-3}$ \citep{LBK2022}. The Galactic magnetar, SGR 1935+2154, is also a repeating source of FRBs \citep{STARE2020,CHIME2020}. Due to its closer distance to us, the typical energy of bursts from SGR 1935+2154 is naturally smaller than for cosmological repeaters. However, a much greater difference between this source and prolific repeaters is its extremely low activity rate above a given energy, $\lambda(>E_{\rm iso})$ which is $\gtrsim 10^5$ times smaller than that of highly active repeaters at a similar $E_{\rm iso}$ \citep{MBSM2020}, signaling that this source is distinct from highly active repeaters.
The volumetric source density of SGR-like sources similar to SGR 1935+2154, is $10^7 \lesssim n_{\rm obs}(\mbox{SGR})\lesssim 3\times 10^8\mbox{ Gpc}^{-3}$ \citep{LBK2022}, i.e. a large factor of order $\sim 10^6$ larger than inferred source densities of prolific repeaters. How can such a huge ratio be explained? Part of this difference may be ascribed to the rarity of the conditions in the underlying magnetar responsible for these events. An important parameter in this regard is the initial magnetic field of the magnetar, $B_0$, that determines the overall energy reservoir and therefore the rate of bursts that can be produced by the NS. That being said the inferred birth (internal) magnetic field of Galactic magnetars is already very large $B_0\approx 10^{14.5}-10^{15.5}\mbox{ G}$ \citep{Dall'Osso2012D,Beniamini2019,Beniamini2024}, and cannot be much larger before the NS becomes unstable \citep{2013MNRAS.433.2445A}. So unless the burst activity scales extremely strongly with $B_0$, this would not be sufficient to explain the source density disparity above. Another option is that there is a large difference in rotation periods between the two classes of sources. This is discussed in detail in \S \ref{sec:ULPM}.
Finally, the association of active repeaters with aligned rotators provides a natural explanation for the extremely low required source density. In this case, the inferred volumetric densities quoted above, should be multiplied by a factor $f_{\rm align}^{-1}f_{\rm b,eff}^{-1}$, where the first factor accounts for the fraction of sources for which $\alpha\lesssim \rho$ and the second for the fraction of sources that have their polar caps aligned with towards us. Put together, this can easily accommodate for a source density increase by several orders of magnitude. 
To demonstrate this, we consider the following, very simplified estimate.
As shown in Eq. \ref{eq:fbeff}, $f_{\rm b,eff}(\alpha<\rho)\approx 4\times 10^{-3} r_6 P_{-1}^{-1}$. 
$f_{\rm align}$ depends on the distribution of $\alpha$ and its potential evolution over the magnetar's age. Based on pulsar observations, \cite{LM1988} estimate that for young pulsars $\alpha$ is approximately uniformly distributed between 0 and $\pi/2$ \footnote{\cite{Tauris1998} find that the pulsar distribution may be intrinsically concentrated towards low inclinations (somewhat increasing $f_{\rm align}$), but also that $P$ is negatively correlated with $\alpha$ (meaning that $f_{\rm b,eff}$ would become smaller). The overall effect on $f_{\rm align}^{-1}f_{\rm b,eff}^{-1}$ is sensitive to the poorly constrained extrapolation of these properties to magnetar field strengths and periods.}. Taking this estimate as a guide also for young magnetars \footnote{If the evolution $\alpha$ is governed by conditions in the NS interior, it may be similar between magnetars and lower magnetic field pulsars.},
$f_{\rm align}\approx 2\rho/\pi \sim 4\times 10^{-2} r_6^{1/2} P_{-1}^{-1/2}$. Put together, we see that with this simple estimate $f_{\rm align}^{-1}f_{\rm b,eff}^{-1}\approx 5\times 10^3 P_{-1}^{3/2} r_6^{-3/2}$ which can accommodate for a large fraction of the observed $\sim 10^5-10^7$ ratio of source densities. As mentioned above, the remaining fraction might be ascribed to a requirement of a sufficiently large magnetic field strength operating in active repeaters.

\item {\bf Repeater to Non-repeater ratio.} The discussion above suggests that, all other things being equal, the observed ratio between highly active repeaters and apparent non-repeaters in a volume-limited sample, should be of the order of $N_{\rm rp}/N_{\rm nrp}\approx f_{\rm align} f_{\rm b,eff}(\alpha<\rho)/f_{\rm b,eff}(\alpha\approx 1)\approx  3\times 10^{-3} r_6^{1}P_{-1}^{-1}$. This can match the observed ratio of one in several hundreds, for reasonable values of the emission height / rotation period.

\end{enumerate}

\subsubsection{Ultra long period magnetars with $\alpha\approx \rho$}
\label{sec:ULPM}
Another suggested explanation for highly prolific repeaters is that they might be associated with a sub-class of ultra-long period magnetars (ULPMs) with rotational periods of order tens of days or more \citep{Beniamini+20}. This possibility naturally fits with the geometrical interpretation suggested in this work. We briefly discuss the merits and drawbacks of aligned ULPMs as active repeaters below.
\begin{enumerate}
\item {\bf Underlying periodicity.} If prolific repeaters are associated with ULPMs, then clearly no short time scale periodicity should be observed. Indeed, this association then naturally explains the observed active phase periodicity of arrival time in R1 and R3 bursts as directly due to rotation. The active phases of R1 and R3 are of order $\sim 0.56, 0.25$ respectively, i.e. not much smaller than unity. Within the ULPM framework, since the polar cap size is very small, $\rho \ll 1$, this demands $\eta=W/2\pi \approx 0.25-0.5$ and therefore $\alpha\approx \rho$, i.e. marginally aligned rotators \citep{Beniamini2023}. 
\item {\bf Energetics.} The consideration here is qualitatively similar to that in \S \ref{sec:Youngmags}, except that due to the much smaller polar cap, the suppression in the required energy is much more prominent. 
\item {\bf PA swing.} Since $t_{\rm br}\ll P$ in this scenario, we would expect no geometrical PAS during bursts. However PAS between bursts observed over a timescale much longer than the period, would typically (depending on the exact configuration of $\alpha, \beta$) be large.
\item {\bf Spectro-temporal properties.} As in \S \ref{sec:Youngmags}, the association of marginally aligned ULPMs with prolific repeaters, leads to fixed conditions of the emitting region during a burst and naturally results in the observed dichotomy between the spectral and temporal properties of CHIME repeaters and non-repeaters.
\item {\bf Source densities.} In the last few years, several long period Galactic objects have been detected in the radio \citep{Hurley-Walker2022,Hurley-Walker2023,caleb2022,Caleb2024}. These newly discovered objects have periods of (so far) up to an hour - much longer than those of previously known Galactic pulsars and magnetars, while showing very similar phenomenology to those objects. Various observational and theoretical considerations, including deep optical limits ruling out a white-dwarf origin or a binary companion, suggest that these are ULPMs \citep{Beniamini2023}. Furthermore, these objects (at least with periods of $\lesssim 1$\, hour) appear to be both old (with ages of $\sim 10^5-10^6$\, yrs) and very abundant (with potentially tens of thousands of such objects or more in the Galaxy). At the same time, the requirement for alignment will severely decrease the number of such objects that would be viewed by us as active repeaters. To quantify this, we would need to first estimate the polar cap size. For a dipole configuration (Eq. \ref{eq:thetapc}), $\rho$ can be as small as $10^{-5}$\,rad. Since this is an extremely small angle, it is unlikely to be the dominant effect dictating the polar cap size. For example, even a moderate outflow from such an object can lead to a significant opening of field lines beyond this angle (indeed in Galactic ULPMs, the  observed duty cycle value $\eta\sim 0.3-5 \%$ suggests that $\rho \gg \rho_{\rm dip}$). For this reason, quantitative beaming estimates should be done with caution in the ULPM scenario.

\item {\bf Repeater to Non-repeater ratio.} The situation here is similar to the one discussed in \S \ref{sec:Youngmags}, except that the expression for $N_{\rm rp}/N_{\rm nrp}$ should be multiplied by the ratio of ULPMs to `normal' magnetars, which as mentioned above may be a factor of $\sim 100$ or more, helping to offset the decrease in the $f_{\rm align}$ and $f_{\rm b.,eff}$ that would be expected in the ULPM scenario.

\end{enumerate}

\subsection{Theoretical considerations} 

It is shown, e.g. \citep{Qu2022}, that it is much easier for waves produced inside a magnetar magnetosphere to escape along open field lines than closed field lines where the angle between the wave-vector and magnetic field direction is expected to be small. Moreover, plasma is likely flowing outward with high Lorentz factor along open field lines making the interaction between the wave and plasma of limited importance as they are both moving at close to the speed of light and therefore have limited region where they can overlap and interact strongly. 

\section{Predictions}
\label{sec:predict}
We list below a few predictions of the model that can be tested with future observations.
\begin{itemize}
    \item Combining the expression of the sweep time (Eq. \ref{eq:tsw}) with the period evolution due to dipole radiation (Eq. \ref{eq:Plim}), we get
    \begin{equation}
    \label{eq:tsw2}
        t_{\rm sw}\approx 9.6 g \left(r_6 B_{d,14.5}\right)^{1/2} \left(\frac{t}{10^4\mbox{ yr}}\right)^{1/4} \mbox{ ms}
    \end{equation}
    where $t\approx 10^4$\,yr is the typical age of a magnetar before its dipole field significantly decays. The typical sweep time is comparable or slightly larger than the typical observed FRB duration. Since it is the minimum between $t_{\rm sw}$ and $t_{\rm br}$\footnote{Another important timescale is the time it takes the peak frequency to scan across the observed band (see last point in \S \ref{sec:repeatersaligned}). If it is shorter than the other two, it will determine the observed burst duration.} that dictates the observed duration (and $t_{\rm sw}\ll t_{\rm br}$ is required for a significant PAS), we expect (primarily) non-repeater bursts with longer observed durations (of order several of milliseconds) to show PASs. These would correspond to a combination of slightly weaker than average magnetic field strength and / or younger age of the underlying magnetars. Furthermore, the fraction of bursts with such PAS should increase with the observed FRB duration. Considering realistic values of $r,B_d,t$ and the week dependence of $t_{\rm sw}$ on those parameters, Eq. \ref{eq:tsw2} suggests also a maximum typical\footnote{This limit regards the typical duration of non-repeater durations, but is not absolute. It is possible to have a magnetar whose polar cap is pointed towards us, but that is intrinsically very inactive and thus appears as a non-repeater which can have a burst duration dictated by the true activity time. Nonetheless, such cases should be the exception rather than the rule.} duration of a non-repeating FRB spike (after accounting for extrinsic effects, such as, e.g., scattering) of order $\sim 30 \mbox{ ms}$.
    \item The conditions discussed above, leading to large PAS, can also lead to periodic bursts, provided that $t_{\rm br}>P$ (and that the additional conditions listed in table \ref{tab:summary} and Eq. \ref{eq:Gammalimpol} are satisfied). This additional requirement means that periodic non-repeaters would be a (potentially rare) sub group of bursts with large PAS. Importantly, this PAS in only observable if the observed duration of the bursts is not dominated by scatter broadening, which would effectively mix at a given observed time, emission from different points in the sweep, with different intrinsic PA (unfortunately, scatter broadening indeed appears to dominate the observed width of individual spikes in FRB 20191221A; \citealt{CHIMEperiodicity}).
    \item Radius-to-frequency mapping, wherein high frequency radio waves originate from lower radii and vice versa, is commonly inferred in emission from pulsars \citep{1970Natur.225..612K,1978ApJ...222.1006C}. Even though the luminosities, and therefore radiation mechanism could be completely different, a qualitatively similar feature could hold true for FRB radiation from magnetars, considering that the frequency of emission likely positively correlates with the underlying magnetic field strength and / or density, and both of these quantities only decrease with radius. This suggests that $t_{\rm sw}$ will be shorter for higher frequencies and this makes it more likely that it will be shorter than $t_{\rm br}$. Overall, the implication is that PAS might be more prominent at higher frequencies.
    \item Considering the measured $P=3.2$\, s of SGR 1935+2154, we expect it to have $t_{\rm sw} \approx 12.5 r_6^{1/2} g$\, ms. If this source ever produces a highly polarized burst with a duration of $\gtrsim 15$\, ms, we expect it to result in a PAS.
    \item While the association of active repeaters with aligned sources and misaligned sources with non-repeaters leads to qualitatively different expectations between the corresponding classes, the distribution of $\alpha/\rho$ in nature, should be continuous. Misaligned sources with $\alpha/\rho\gtrsim 1$ are more likely to be detected as repeaters than sources with $\alpha/\rho \gg 1$. Such marginally misaligned NSs should have lower observed repetition rates and larger geometrical PAS than very active repeaters. All other things being equal, we might therefore expect to see an anti-correlation between geometrical PAS (i.e. those that follow an S-curve swing) and repetition rate at a given energy. In addition, the underlying geometry of the rotating polar cap scenario can be translated (under certain assumptions) to a distribution of repetition numbers of FRBs detected by a blind radio survey. In appendix \ref{sec:statrepetitionrates} we present such a distribution and show that results from the 1st CHIME catalog are consistent with this scenario. We also outline how future observations can be used to provide critical tests of this picture. In particular, we show that quasi periodic signatures in the arrival time data of bursts from a repeater seen by a large field of view survey would require a much more sensitive survey than CHIME and would likely be dominated by a sub-population of sources that are much more active than typical ones.
\end{itemize}

\section{Conclusions}
\label{sec:conclusion}

It is interesting to note that the conclusion that FRB 20221022A is exhibiting pulsar-like behavior powered by magnetic energy dissipation rather than pulsar spin-down is similar to the inference made in the case of the long-duration FRB 20191221A \citep{BK2023}. Moreover, the radio data for both these bursts also suggest that burst properties are consistent with the radiation being produced inside the magnetar magnetosphere in the polar cap region of open field lines and require similar underlying conditions. 
However, the specific arguments differ for the two bursts.

An important question, thus far unanswered, can finally be probed using the polarization data for FRB 20221022A. The question is: whether the duration of the FRB reflects the duration of the NS outburst or if it is the duration of a narrow beam that sweeps past us. The data for FRB 20221022A suggests that, for this burst, it is the latter. The reason being that if the outburst lasted for a time much smaller than the beam-passage time, the PA would have changed only by a small amount and not by the 120$^{\circ}$ as observed. It would be quite a coincidence for the outburst to have lasted for a time just equal to the beam transit time since the two are independent variables. Thus, most likely, the burst lasted for a time longer than the observed 2.5 ms duration. At the same time, the duration of activity likely lasted less than $\sim 130$\,ms, which is the inferred rotational period in this picture, otherwise it would have exhibited periodic modulations, as in the case of FRB 20191221A. Generalizing, our model suggests that (ignoring fine tuned geometries) the relation between activity time, $t_{\rm br}$, the polar cap sweep time, $t_{\rm sw}$, and the NS's rotational period, $P$, drives different observed phenomenology. When $t_{\rm br}<t_{\rm sw}$, bursts appear as single spikes with no significant PAS. For $t_{\rm sw}<t_{\rm br}<P$ we will observe a single spike with potentially large PAS. Finally, for $t_{\rm br}>P$ bursts will have periodic modulations, each potentially showing large PAS. As more bursts can be grouped into these categories, we will be able to constrain the range of activity times of the source and constrain the underlying explosion mechanism.

Under the conservative assumption that entire region of the polar cap is unlikely to be emitting coherently, the data also allows us to place an interesting lower limit on the Lorentz factor, $\Gamma$, of particles in the radio beam, assuming that the high $\Gamma$ is responsible for the observer seeing only a small part of the beam at a time\footnote{If the observer were to receive photons from the entire beam at one time, then little polarization swing will be observed.}. Since the degree of linear polarization observed for FRB 20221022A is $95\pm3$\%, and the full swing of the PA is 120$^{\circ}$, at most 10\% of the area of the radio beam should be visible to the observer at any time (see Eq. \ref{eq:PoldegGamma}). Since, as argued in \S \ref{sec:evidencePC}, the duty cycle for this burst is $\sim 1$\%, that means that the particle LF should be larger than $\Gamma \gtrsim 100$.

We posit here that prolific FRB repeaters are associated with aligned (or marginally aligned) rotators, for which the angle between the spin and magnetic axes is comparable or smaller than the beam size associated with the polar cap. This simple geometrical difference between prolific repeaters and apparent non-repeaters naturally explains many of the features of highly active repeating FRB sources.
We consider two variations on this idea: one, motivated by the large persistent luminosity and rotation measures associated with some active repeaters \citep{Hilmarsson2021a,Chatterjee+17}, which involves young and particularly strong $B$ field magnetars and the other, motivated primarily by very long timescale periodic activity windows, which involves ultra-long period magnetars \citep{Beniamini+20}. We focus in this summary on the first scenario (for which the underlying objects are better understood and numerical estimates are more reliable).
Evidence supporting the first scenario includes: (i) the lack of observed short timescale periodicity in arrival time of bursts, the reason being that the polar cap beam remains within the LOS at all times instead of periodically sweeping in and out. (ii) The frequent bursts and the overall large isotropic equivalent energy release inferred for FRBs with high repetition-rates are consequences of the bursts occurring in the polar cap region. All of the bursts produced in these objects are visible to the observer due to the near alignment of the magnetic and spin axes with the observer's line of sight (LoS). The beam-corrected energy budget of the source is smaller by a factor of $\sim \rho\approx 0.07(0.1\mbox{ s}/P)^{1/2}$  compared to misaligned rotators, and by $\rho^2/2\approx 2\times 10^{-4} (0.1\mbox{ s}/P)$ compared to a NS with completely randomly distributed bursts on its surface. The observed rate is also larger (relative to a misaligned rotator or NS with random bursting sites) by the inverse of the same factor. As the burst rate ratio between repeaters and non-repeaters is larger than can be accounted for by this geometrical factor alone \footnote{Conversely, in the ULPM scenario, geometrical factors alone can be large enough to account for the burst rate ratio between active repeaters and non-repeaters.}, an additional large factor between the rates must be attributed to a distinct property such as the different magnetic field strengths of the underlying magnetars\footnote{Indeed the two requirements might be closely connected: the large $B$ for active repeaters might explain why the magnetic and spin axes align more quickly than in `normal' magnetars.}  (iii)
No geometrical PA swing during bursts \footnote{A few rare bursts from prolific repeaters have been reported to show large PA jumps \citep{Niu2024}. Their timescale and shape suggest their origin is unlikely to be geometrical as discussed in this work, and they might instead be related to the emission mechanism or a propagation effect. Empirically, one can test whether or not a rotating vector model with consistent underlying parameters can account for these jumps in order to confirm or rule out a geometrical origin.} and little swing between bursts (since the PA is determined by the projection of the magnetic axis on the sky plane). (iv) The spectrally narrow and temporally longer nature of repeater as opposed to non-repeater bursts. This is because for aligned rotators, the region of the polar cap visible to the observer barely changes with time relative to the NS surface, and the physical conditions in the emitting region change less as compared to when the line of sight sweeps through a large segment of the polar cap.
(v) The observed detection rate of highly prolific repeaters compared to non-repeaters is approximately given by the fractional solid angle of the polar cap, i.e. $\sim 2\rho^2/\pi \approx 3\times 10^{-3} (0.1\mbox{ s}/P)$, which is close to the observed ratio for large field of view surveys such as CHIME \citep{CHIME_1st_cat}. (vi) The much smaller source densities inferred for active repeaters compared to those of normal Galactic magnetars follows directly from our model as active repeaters require `double coincidence', i.e. magnetic and spin axes are closely aligned and that both are also closely aligned with the observed LOS. This suggests that the fraction of magnetars that are repeating FRBs should be of the order of or smaller than $\sim \rho^3/\pi \sim 1.8\times 10^{-4} (0.1\mbox{ s}/P)^{3/2}$. The observed fraction is between $\sim 10^{-7}-10^{-5}$ \citep{LBK2022}; as above the difference between our estimate and the observed rate can be naturally accounted for by the possibility that active repeaters are likely magnetars with stronger magnetic field strengths than Galactic magnetars.

The picture presented in this work leads to several predictions that could be tested by future data: (i) the typical maximum duration of a pulse for non-repeating FRBs should be on the order of 30 milliseconds (this is due to the weak dependence of $t_{\rm sw}$ on the emission height as well as the underlying magnetar's age, field strength). (ii) Considering its observed spin period, if SGR 1935+2154 ever produces a highly polarized radio burst of duration $\gtrsim 15\mbox{ ms}$ (i.e. a duration longer than its $t_{\rm sw}$, estimated from its observed $P$), it should result in a significant PA swing. (iii) Bursts with large PA swings should have longer-than-average durations, and those with periodic pulses (such as FRB 20191221A) should have even longer durations still. The latter are also generally expected to be a subclass of the former\footnote{Provided that individual spikes are not significantly scatter broadened, which will mix radiation produced at different points during the sweep, and wash out the PA swing evolution.} (this is because the observed duration of a spike is the minimum between $t_{\rm sw}$ and the intrinsic duration, and because the latter needs to be $>P>t_{\rm sw}$ for us to observe periodic spikes). (iv) PA swing may become more prominent at high frequencies, which likely originate from lower emission heights and correspond to shorter $t_{\rm sw}$. (v) The model also predicts an anti-correlation (with large scatter) between the degree of geometrical PA swing (which decreases with $\rho/\alpha$) and the activity rate at a given energy (which increases with $\rho/\alpha$) and will eventually enable the detection of quasi-periodic signatures in the arrival time distribution of bursts from moderately active, slightly misaligned, sources.

\appendix
\section{FRB energy Source}
\label{sec:energy}
\subsection{Non-repeaters}
The isotropic radio luminosities of FRBs 20221022A and 20191221A ($L_{\rm iso}=1.05\times 10^{41}\mbox{erg s}^{-1}$ and $L_{\rm iso}=2\times 10^{41}\mbox{erg s}^{-1}$ respectively), together with the estimate of $P$ based on the observed PA swing / individual pulse durations, strongly constrain the burst energy source. 

{\bf Spindown powered.} We consider first the possibility that the FRB is powered by spindown. Denoting the radio efficiency by $\epsilon_r$  we can establish a relationship between the spin-down luminosity ($L_{\rm SD}$), the beaming-corrected radio luminosity ($L_{\rm rad}$), and the observed isotropic equivalent radio luminosity ($L_{\rm iso}$): $L_{\rm SD}=\epsilon_r^{-1}L_{\rm rad}=\epsilon_r^{-1}\rho^2L_{\rm iso}/4$. Using this relation, we can determine the required dipolar magnetic field at the NS surface as 
\begin{eqnarray}
\label{eq:BSD}
   && B_{\rm d}\!=\!2\!\times \!10^{16} P_0^{3/2} L_{\rm iso,41}^{1/2} r_6^{1/2} \epsilon_{r,-3}^{-1/2} \mbox{ G} \\ 
   && =10^{15} \left(\frac{t_{\rm sw}}{2.5\mbox{ ms}}\right)^3 L_{\rm iso,41}^{1/2} g^{-3} r_6^{-1} \epsilon_{r,-3}^{-1/2} \mbox{ G} \nonumber
\end{eqnarray}
where 
in the last equation we have used the estimate of $P$ from Eq. \ref{eq:P221022}. This demonstrates that a magnetar strength (or even higher) magnetic field strength is required. Since the rate of magnetic energy decay in such objects is much higher than the spindown power \citep{TD1996}, this already argues against spindown as the FRBs' energy source. An independent argument comes from considering the spindown age. Using the relation between $B_{\rm d}$ and $P$ from Eq. \ref{eq:BSD}, the implied spindown time due to dipole radiation is
\begin{equation}
    t_{\rm SD}=0.6 g^2 \left(\frac{t_{\rm sw}}{2.5\mbox{ ms}}\right)^{-2} L_{\rm iso,41}^{-1} \epsilon_{\rm r,-3}\mbox{ yr}
\end{equation}
independent of the emission radius. For 0.1\% efficiency, which is a highly optimistic value for $\epsilon_{\rm r}$ (see \citealt{Szary2014}), the spin-down age of the NS is less than a year, which is highly unlikely.
The reason is that the ionized debris from the formation of such a young system would contribute $>10$ cm$^{-3}$ pc to the observed DM\footnote{The ejecta is likely fully ionized due to UV \& X-ray emission from the NS spin-down and cooling.}, even when the total mass ejected in the formation of the NS is $\sim0.1 M_\odot$ (i.e. a rare magnetar formed by a binary NS merger), and the ejecta speed is 0.1 c. This DM is larger than the strong upper limit placed by observations of FRB 20221022A on DM$_{\rm host}\lesssim 14\mbox{pc cm}^{-3}$ \citep{mckinven2025}. Moreover, young nebulae around NSs, fed by the magnetar wind, and the interaction of these nebulae with the ISM should give rise to persistent bright radio emissions, which does not appear to be the case with the exception of three cases out of about 10$^3$ FRBs thus far observed.

{\bf Magnetically powered.} An alternative explanation is that FRBs 20221022A, 20191221A are powered by the dissipation of magnetic field energy rather than the rotational energy.

As above we estimate the spindown time in this scenario,
\begin{equation}
    t_{\rm SD}=70 B_{\rm d,14}^{-2}\left(\frac{t_{\rm sw}}{2.5\mbox{ ms}}\right)^{4} g^{-4} r_{6}^{-2} \mbox{ yr}
\end{equation}

We see that older NS ages are permitted in this interpretation as compared to the spindown powered case. Under this interpretation, the rate of energy loss in the radio band is $L_{\rm rad}=8.6\times 10^{37} r_6^2g^2 (2.5 \mbox{ ms} /t_{\rm sw})^2\mbox{erg s}^{-1}$. This is well within the luminosity of short X-ray bursts from magnetars, which can have peak luminosities as large as $10^{43}\mbox{ erg s}^{-1}$ and are well accepted as being magnetically powered \citep{KaspiBeloborodov2017}.

\subsection{Repeaters}
\label{sec:energyrepeat}
The observed energy release in some FRB repeaters is $E_{\rm ob,iso}\sim 1-10\times 10^{41}\mbox{ erg}$. A good case is that of the famous repeater R1 which released $E_{\rm ob,iso}=3.4\times 10^{41}\mbox{ erg}$ during $t_{\rm ob}=59.5$\,hrs of observation by FAST \citep{2021Natur.598..267L}. Considering that R1 has a duty cycle of $56\%$, with a period of $\sim 160$\,d \citep{Rajwade2020},  much longer than the duration of the FAST observation, the average isotropic equivalent luminosity of R1 can be approximately estimated as $L_{\rm ob,iso}=E_{\rm ob,iso}\times 0.56/t_{\rm ob}\approx 9\times 10^{35}\mbox{erg s}^{-1}$. We denote by $t_{\rm act}$ the duration over which R1 produces bursts at a comparable rate to the one observed today.
 Since R1 has remained highly active in the 12 years since it was first detected, it is clear that $t_{\rm act}\!>\!12$\,yrs, and considering the lack of evolution in burst energetics during this interval, it is likely that $t_{\rm act}$ is quite a bit longer. 
The total energy release by the underlying magnetar associated with its FRB activity, is
\begin{equation}
\label{eq:Erel}
    E_{\rm rel,R1}=8.4\times 10^{44}\left(\frac{f_{\rm b,eff}}{\epsilon_r}\right)\left(\frac{t_{\rm act}}{30\mbox{ yr}}\right)\mbox{ erg}
\end{equation}
where $f_{\rm b,eff}$ is the effective beaming (i.e. the fraction of energy emitted by a source that is detectable by a given observer with an infinitely sensitive telescope). The latter can be much larger than the actual beaming factor associated with any single burst, $f_b$. We consider three illustrative examples to demonstrate this: (i) if the direction of bursts relative to the observer is completely random, then each burst's isotropic energy is larger by $f_b^{-1}$ than its true energy, but only a fraction $f_b$ of produced bursts are observed, such that $f_{\rm b,eff}=f_b^{-1}\times f_{\rm b}=1$. Considering a realistic radio efficiency, $\epsilon_{\rm r}\lesssim 10^{-3}$ (and perhaps even much smaller), the fact that $f_{\rm b,eff}=1$, combined with Eq. \ref{eq:Erel} strongly constrains the possibility of random bursts on the surface based on energetic considerations. (ii) If the bursts are mostly produced in the magnetic polar cap, by a highly relativistic plasma flowing along the field lines, and the magnetic pole is rotating around the spin-axis, then the fraction of observed bursts is increased to $f_b f_{\rm sw}^{-1}$ where $f_{\rm sw}\approx  \rho\sin \alpha  <1$ is the fractional area swept by the two (for a dipole geometry) burst-emitting polar caps, while the ratio between observed and isotropic burst energy remains unchanged as compared with (i). 
(iii) If the bursts are always produced towards the observer, i.e. when $\alpha\ll \rho$, then the fraction of observed bursts increases to $f_{\rm b}f_{\rm pc}^{-1}$ where $f_{\rm pc}\approx \rho^2/2$ is the fractional area covered by the emitting polar caps. 
To summarize the effective beaming is given by
\begin{eqnarray}
\label{eq:fbeff}
    f_{\rm b,eff}\!=\!\left\{ \!\begin{array}{ll} \! 1 & \mbox{random loc.}   \\ 
      \! f_{\rm sw}\!=\!0.02\sin \alpha (r_{\rm 6}/P_0)^{1/2} & \mbox{pc sweep }(\alpha\!\gtrsim\!\rho) \\ 
      \! f_{\rm pc}\!=2\times 10^{-4}(r_{6}/P_0) & \mbox{fixed pc }(\alpha\!\ll \!\rho). 
\end{array} \right.   
\end{eqnarray}

Considering the energy source to be magnetic, we can use Eq. \ref{eq:Erel} to estimate the minimum required magnetic field such that $E_{\rm B}>E_{\rm rel}$,
\begin{equation}
    B>2\times 10^{15}f_{\rm b,eff,-2}^{1/2} \epsilon_{\rm r,-5}^{-1/2} \left(\frac{t_{\rm act}}{30 \mbox{ yr}}\right)^{1/2} \mbox{ G}
\end{equation}
Taking the surface dipole field to be a fraction $f_{\rm d}$ of the internal field, we use the period evolution due to dipole spindown and find a lower limit on the current value of $P$,
\begin{equation}
\label{eq:Plim}
P\geq \sqrt{\frac{2\pi}{3}}\frac{f_{\rm d}B_0 R^3 \tau^{1/2}}{I^{1/2}c^{3/2}}=50 f_{\rm d,-1} f_{\rm b,eff,-2}^{1/2} \epsilon_{\rm r,-5}^{-1/2} \left(\frac{t}{30 \mbox{ yr}}\right)  f_{\rm act}^{-1/2} \mbox{ ms}.
\end{equation}
where $f_{\rm act}=t_{\rm act}/t\leq 1$ is the fraction of its lifetime ($t$) during which the magnetar is actively producing radio bursts. Eq. \ref{eq:Plim} shows that in a magnetic powered scenario, the period of the magnetar powering R1 should be at least of the order of tens of ms. However, \cite{2021Natur.598..267L} have shown that a periodicity with such an underlying period (and $\lesssim 10$\,s, as observed for known Galactic magnetars) can be ruled out by considering the arrival time data of bursts from R1. As described in \ref{sec:repeatersaligned}, this requires a mechanism to `hide' the underlying period - a geometrical alignment between the magnetic and rotation axis is a natural way to do that, while also significantly relaxing the energetic requirements (Eq. \ref{eq:fbeff}).

\section{Beaming and Polarization}
\label{sec:beamingandpolnumerical}
\begin{figure}
\centering
\includegraphics[scale=0.12]{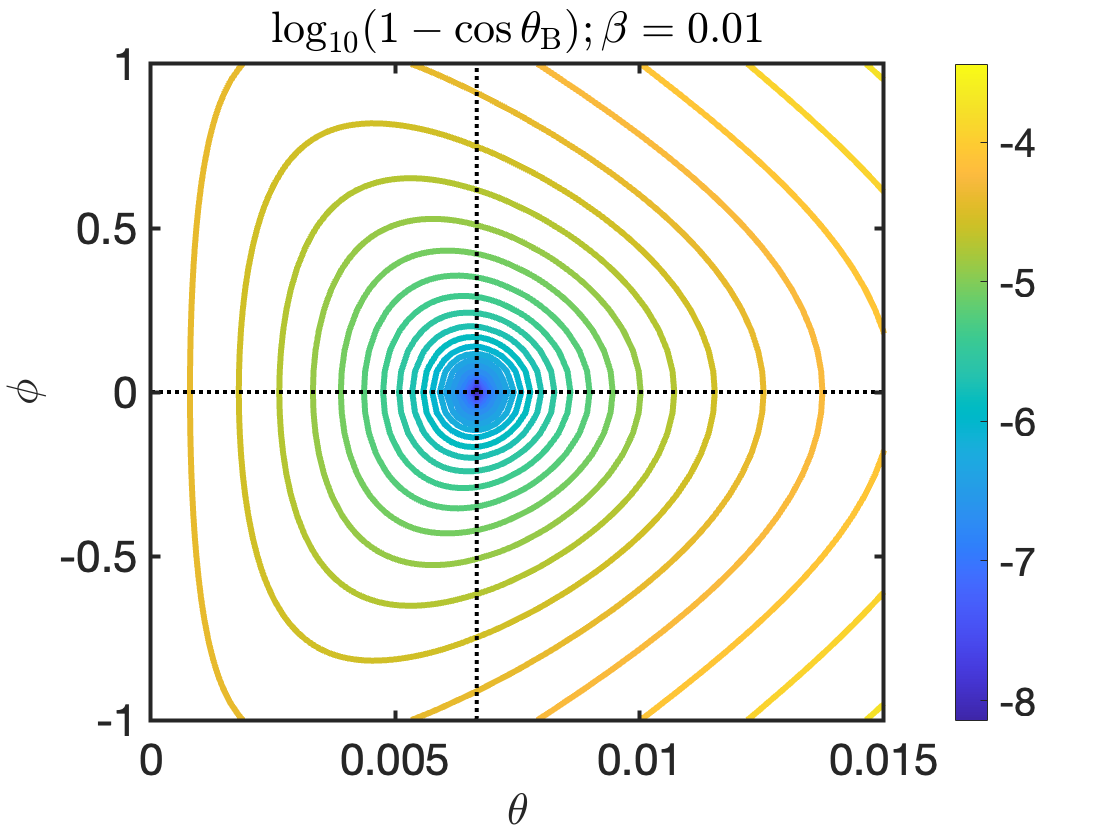}
\includegraphics[scale=0.12]{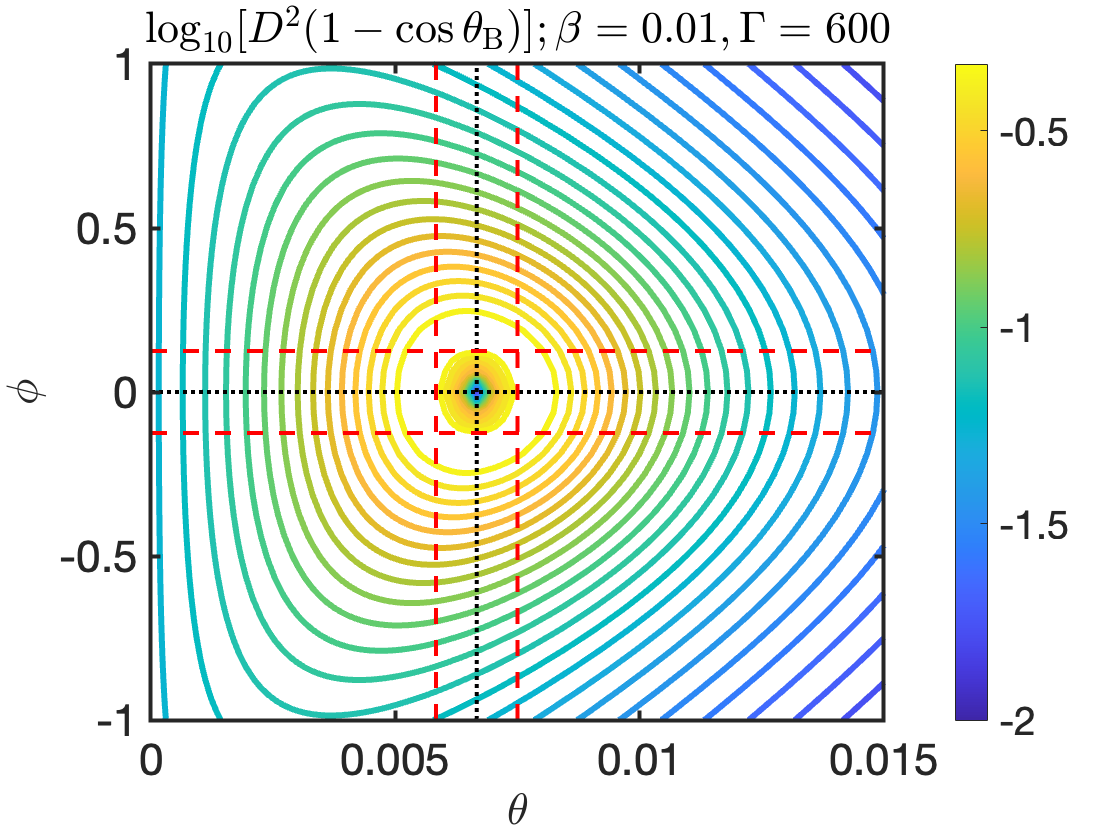}
\includegraphics[scale=0.12]{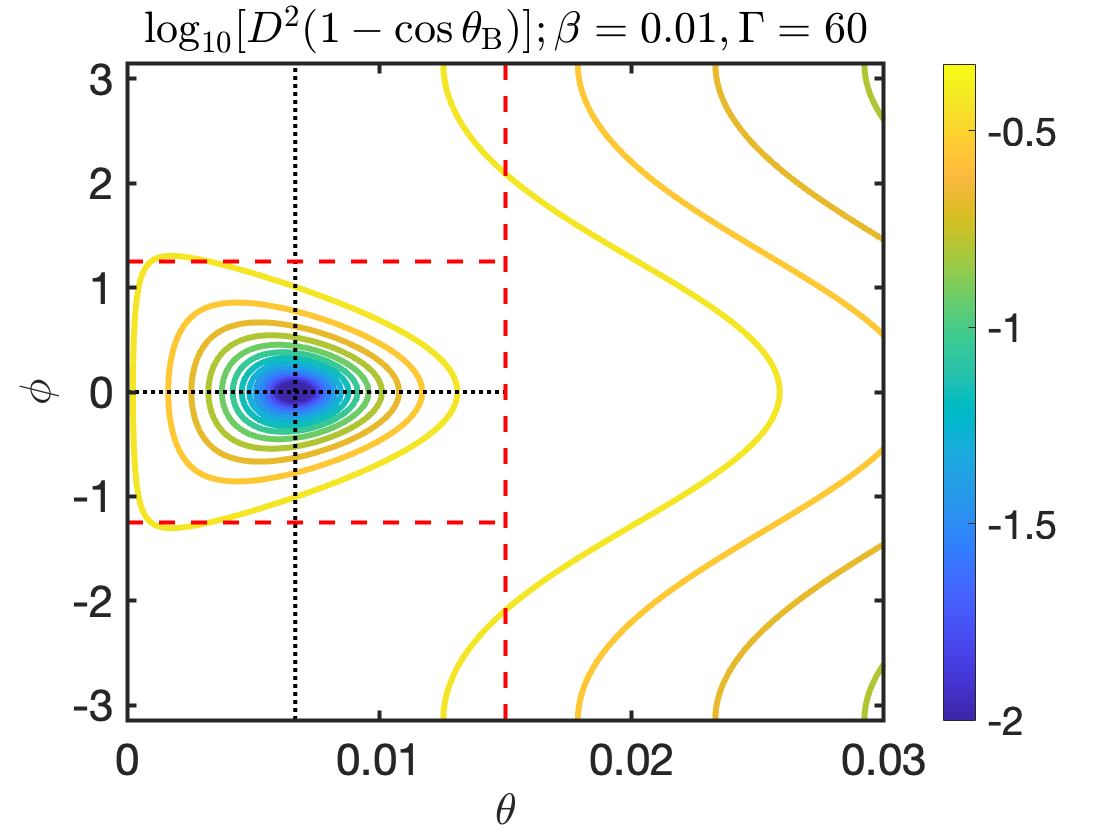}\\
\includegraphics[scale=0.12]{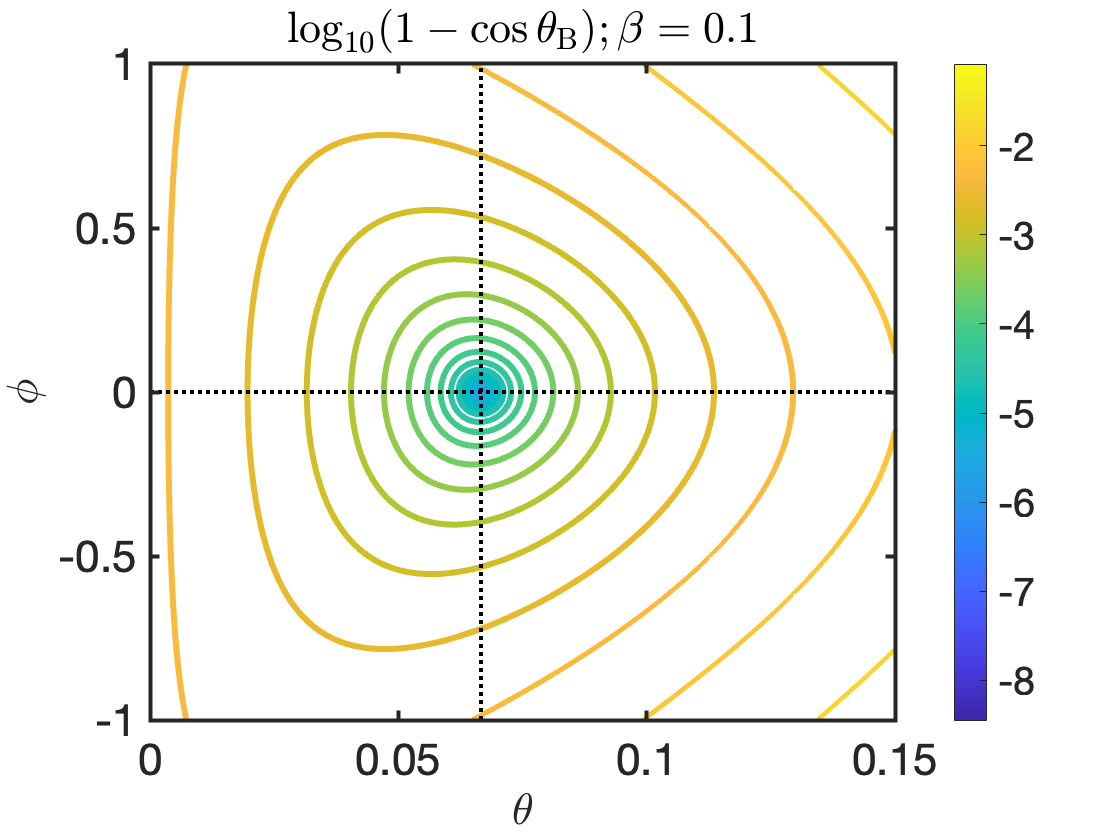}
\includegraphics[scale=0.12]{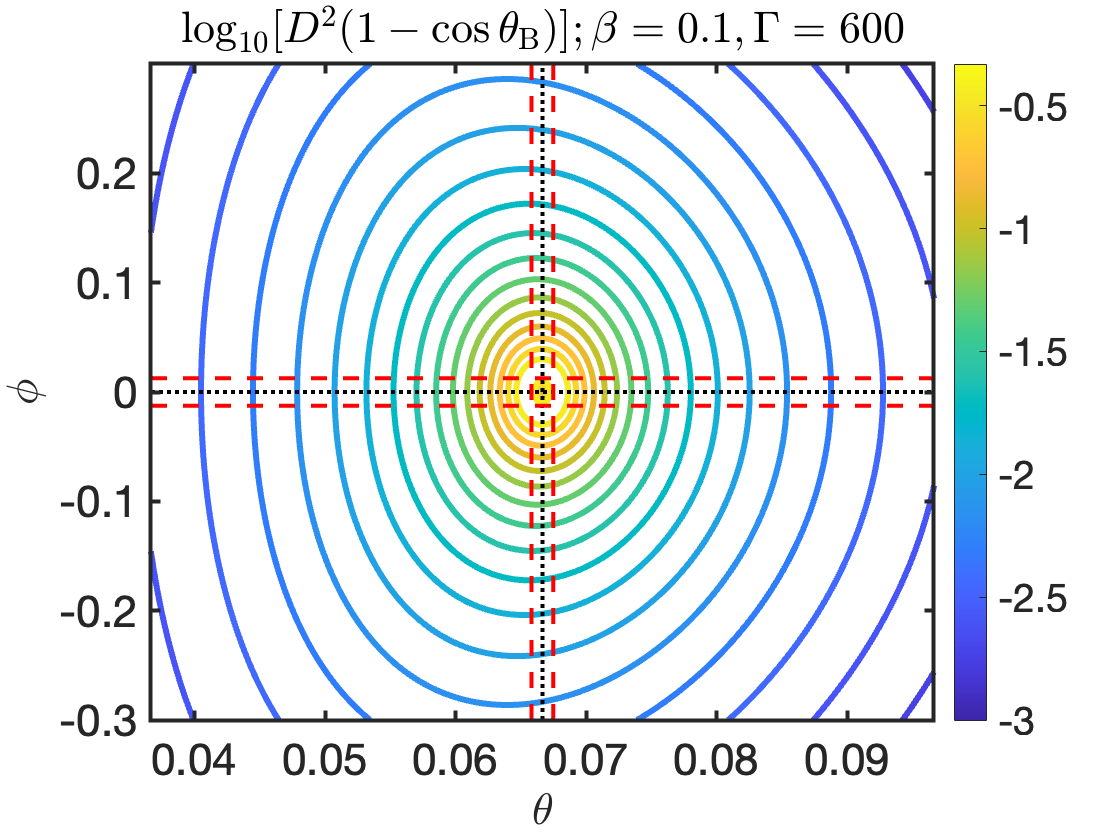}
\includegraphics[scale=0.12]{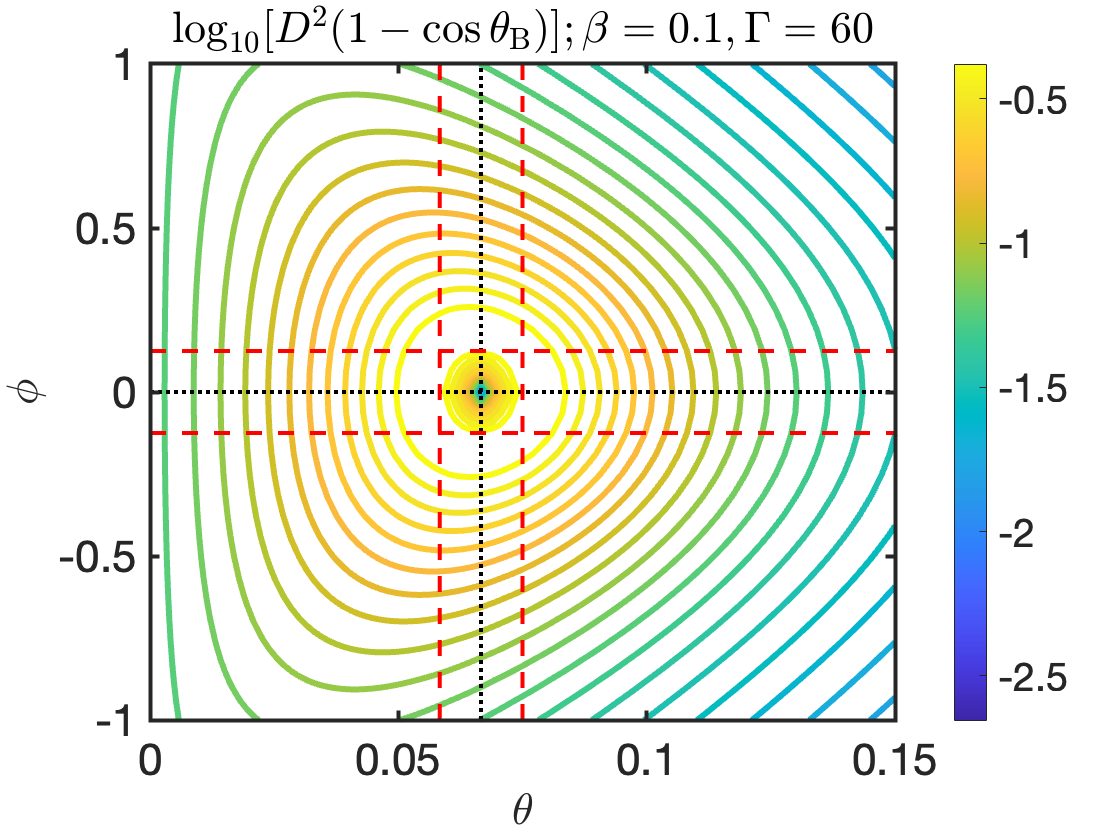}
\caption{Left: Contours of $\log_{10}(1-\cos \theta_{B})=\log_{10}(1-\hat{n}_{\rm obs}\cdot\hat{B}(\theta,\phi))$ as a function of $\theta,\phi$. Black dotted lines depict the emission center point, $\theta_0,\phi_0$ (for which $\vec{B}(\theta_0,\phi_0)\parallel \hat{n}_{\rm obs}$), reproducing Eq. \ref{eq:emcenter}. The values increase rapidly for larger $|\phi|,|\Delta \theta|=|\theta-\theta_0|$ as shown by Eq. \ref{eq:SeriescosthetaB}. Center and right: Contours of the same quantity with appropriate weighting for its contribution to the Stokes parameters in terms of the Doppler factor, $\log_{10}[\mathcal{D}^2(1-\cos \theta_{B})]$. The results recover the analytic result that the contribution is maximized for $|\phi_{\Gamma}|\lesssim 1/(2\theta_0 \Gamma_0)$ and $|\Delta \theta|<(1/2\Gamma)$ which are denoted by dashed red horizontal and vertical lines. Assuming emission takes place within a large fraction of the polar cap, large polarization requires that the effective contribution comes from a small region in this plane - this clearly shows that $\Gamma\gg \theta_0^{-1}$ is necessary to explain high polarization (with more precise limits given by Eq. \ref{eq:PoldegGamma}).}
\label{fig:beamandDoppler}
\end{figure}

We present in figure \ref{fig:beamandDoppler} the values of $(1-\cos \theta_{B})$ and $\mathcal{D}^2(1-\cos \theta_{B})$ which appear in our calculation of the polarization degree described analytically  in \S \ref{sec:poldeg}, as a function of $\theta,\phi$. The values here are evaluated from the full numerical expression, with no small angle approximations. The results match very well with the analytic approximations as shown by comparing the region within the red dashed lines and the highest level contours. The approximation gets somewhat less accurate when $\Gamma\lesssim \theta_0^{-1}$, but this is exactly because at that limit a large range of $\phi$ values contribute significantly to the polarization and as a result the polarization degree is low. In other words, when the polarization is of order unity, it is guaranteed that contributions arise from $|\phi|\ll 1$.

\section{Statistical distribution of repetition rates}
\label{sec:statrepetitionrates}
The geometric picture presented in this work can be used to calculate the distribution of repetition rates as observed by a radio sky survey. On the data side, the most appropriate comparison is with CHIME, which scans the sky and is not targeting specific sources. We will use the repetitions rates reported in the first CHIME catalog \citep{CHIME_1st_cat}.

A full treatment of the detectable distribution of bursts and their comparison with the observed sample, depends on many details that are not apriori determined by the model such as possible correlations between magnetic inclination and intrinsic rate (both likely depend on the magnetic field strength) or correlations between the degree of beaming of individual bursts and their isotropic equivalent radiated energy. Furthermore, as only 16/492 sources detected in the first CHIME catalog were seen to repeat within the considered time range, we will be dealing with small number statistics of repeating sources. As such, the data at present is not constraining enough to probe some of the potential complexities described above. In light of this, we present here the simplest possible realization of the rotating beam scenario. 

We consider magnetars with different magnetic inclinations, characterized by a distribution $dN/d\alpha$. For simplicity and concreteness, we take $dN/d\alpha=const$ (motivated by young pulsar observations, see \S \ref{sec:Youngmags}). We stress that this distribution is not observationally constrained for magnetars and that the results can change significantly based on this choice.
We assign each source a value of $\alpha$ based on this distribution and then draw the viewing angle, $\theta_{\rm obs}$ of the observer relative to $\vec{\Omega}$ from an isotropic distribution, $dN/d\cos \theta_{\rm obs}=const$. Each such source is assumed to have the same intrinsic burst rate. The relative fraction of time that each point on the sky has been observed is taken according to the CHIME distribution reported in the first catalog \citep{CHIME_1st_cat,Josephy2021}. With those assumptions in place, there is a typical number of bursts per source, $\lambda$ produced while the source was in the telescope's field of view. We then consider the bursts to be produced only within the polar caps (each illuminating a size comparable to the full polar cap and having $\Gamma\gg \theta_{\rm pc}^{-1}$, see \S \ref{sec:poldeg}). Since the polar caps are generally rotating relative to the line of sight, most bursts will be missed by the observer (see \S \ref{sec:energyrepeat}). The rotation phase at which bursts are produced is reproduced by taking $\Omega t$ to be distributed uniformly between 0 and $2\pi$. Since $\Gamma\gg \theta_{\rm pc}^{-1}$, we consider bursts to be observable if the observer lies within the polar cap at the time of a burst. We continue generating sources and bursts in this manner using a Monte Carlo approach until we obtain a sample of 492 detected sources, matching the sample size of the first CHIME catalog.

\begin{figure}
\centering
\includegraphics[scale=0.12]{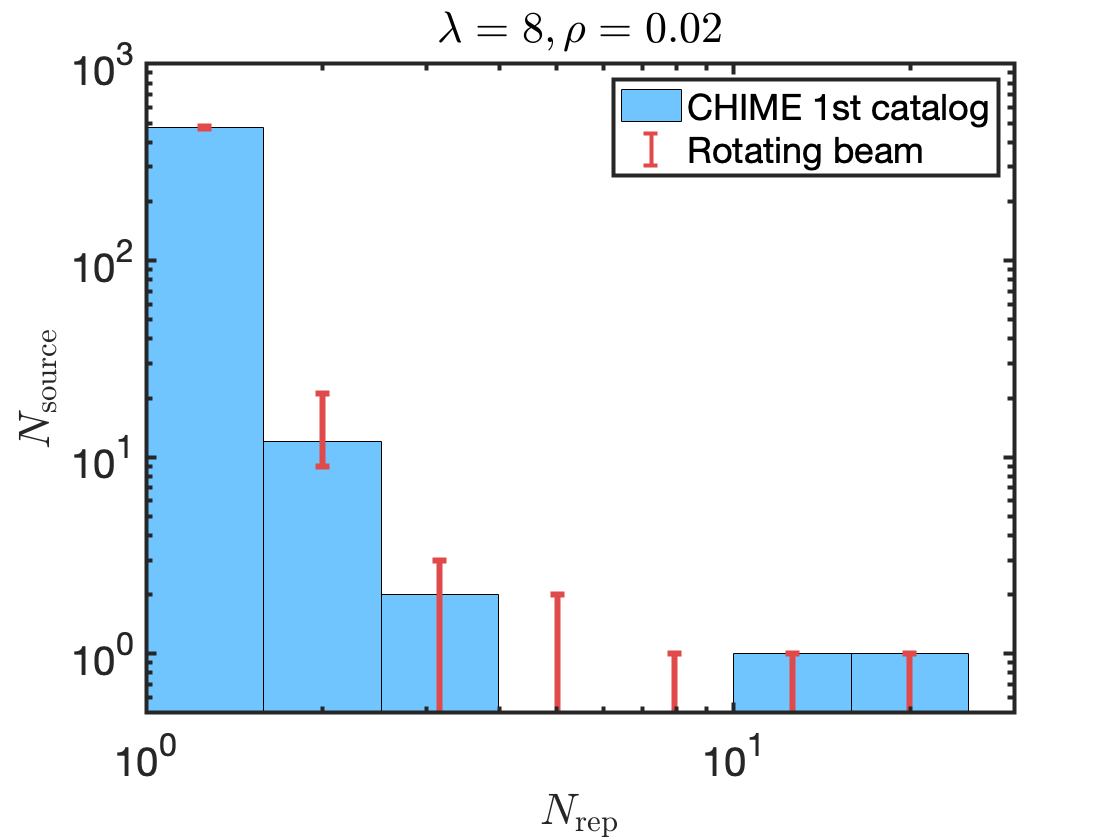}
\includegraphics[scale=0.12]{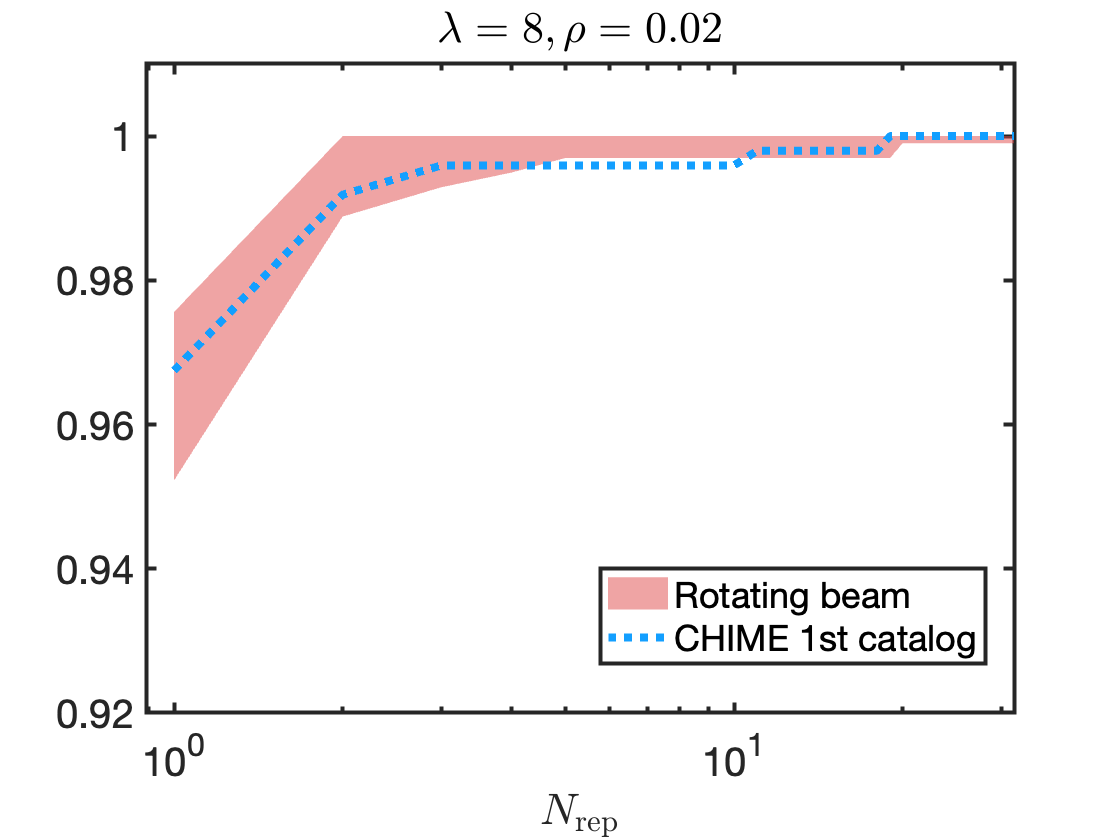}
\includegraphics[scale=0.12]{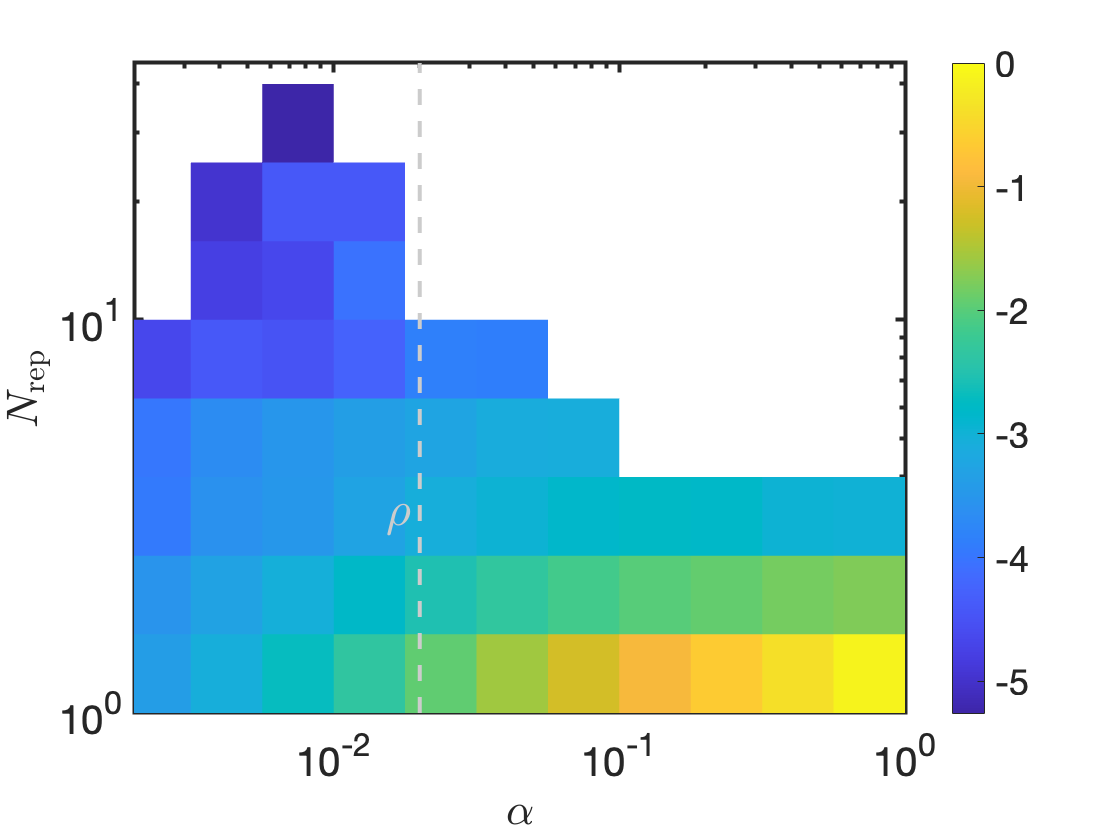}\\
\caption{Left: Histogram of observed number of repetitions per source. Thick bars depict the result for the first CHIME catalog and error-bars depict the $2\sigma$ range for the rotating beam scenario described in \S \ref{sec:statrepetitionrates} for $\lambda=8,\rho=0.02$. Center: The cumulative distribution corresponding to the PDF on the left panel. The two distributions are statistically consistent. Right: Normalized number of observed sources within given logarithmic intervals of $\alpha$ and $N_{\rm rep}$ for the same setup as shown in the left panel. A large repetition number is correlated with $\alpha<\rho$ (left of the dashed grey line) and vice versa.}
\label{fig:repetitions}
\end{figure}

We repeat the process above multiple times and calculate 2$\sigma$ confidence intervals for the cumulative distribution of detected repetitions per source, $N_{\rm rep}$. This distribution is compared with the observed sample from CHIME in Fig. \ref{fig:repetitions}. 
This simplified version of the rotating beam scenario has only two parameters. The first, $\rho$, is confined to a relatively narrow range based on physical considerations (see Eq. \ref{eq:thetapc}).
The second parameter, $\lambda$, depends on the effective observed time per point on the sky and the intrinsic burst production rate per source (at a large enough energy to be detectable). It can be constrained by the requirement that the observed ratio of sources observed twice and once is reproduced. Using Poisson statistics the latter is $\frac{N_2^{\rm obs}}{N_1^{\rm obs}}=\frac{\lambda}{2}P_{\rm ob}(\rho,\alpha,\beta)\approx \frac{12}{476}$ where $P_{\rm ob}(\rho,\alpha,\beta)$ is the probability that the bursts, produced during a random phase of the magnetar's rotation, are pointed towards the observer. For $\alpha\sim 1, \rho\ll1,0<\beta<\rho$, the latter is of order $\rho/\pi$. As a result we get $\lambda\approx 8(0.02/\rho)$.
As evident from the figure (and corroborated using a KS test), even this simplest possible variant of the geometric rotating beam scenario (with physically reasonable values of $\rho, \lambda$) is consistent with the available data. As more data gathers, it will be possible to test this scenario in more detail. At that stage one should account for the underlying intrinsic distributions of periods and burst rates as well as for correlations between the model parameters as discussed above. In addition, on the observational side, one should account for the fraction of observable bursts considering their fluence, frequency dependence, etc.

It is interesting to consider whether it should be possible to observe periodicity in the arrival time of bursts from sources seen multiple times by the survey. 
To see such a signal, a source would need to be detected $\gtrsim 5$ times at the very least within a time of say $T_{P}\approx 5P$ (the latter is required because for slightly misaligned sources, which as discussed below, are the most promising for arrival time periodicity detection, the observable phase is a large fraction of the period, meaning that if the observed signals are sparsely populated relative to the period, the periodic nature of the arrival time data will be erased). From $\lambda$ and the mean time a random point on the sky was observed by CHIME during the period of time culminating in the 1st catalog, $T_{\rm chime}\approx 20\mbox{ hr}$, the mean number of bursts emitted by the source during this interval is $\lambda_{T_P}\equiv \lambda T_{P}/T_{\rm CHIME}\approx 10^{-4}$ (for a typical $P\approx 200\mbox{ ms}$). The probability to detect all these bursts is much lower still, because most bursts will be pointed away from the observer. The best case scenario is to have  $\rho
\lesssim \alpha \lesssim 2 \rho$ (at lower $\alpha$ the source becomes always `on axis' and the imprint of periodicity is lost, while at much larger $\alpha$ the vast majority of bursts produced by the source are pointed away from the observer). In this case the probability to observe all five bursts is of order $2^{-5}$, and the probability of a random source having this orientation is $\approx 2\rho/\pi$. Overall, we see that the probability of detecting a source with periodic arrival time signal is $P_{\rm per}\approx \frac{1}{2^{5}}\frac{2\rho}{\pi}\lambda_{T_P}\approx 4\times 10^{-8}$.
Considering that CHIME has detected 492 sources within the duration of the first catalog, it is no surprise that none of these sources show periodicity in the arrival time data\footnote{The periodic signal found within the lightcurve of FRB 20191221A discussed in \S \ref{sec:evidencePC} is a distinct phenomenon, as it does not require multiple independent triggers and relies instead on a long enough activity time of the source, allowing for the periodic modulation to be detectable within it. Depending on the as of yet poorly determined intrinsic activity time distribution, periodicity may be easier to detect via such a channel.}. This conclusion may be somewhat relaxed if there is a sub-population of sources that are orders of magnitude more active than the typical source currently detected by CHIME. In addition, a more sensitive survey will enable to probe lower intrinsic energies (where the repetition rate is likely much higher). This (together with improvements in the field of view) could make the detection of periodicity in arrival time data of FRBs detected by a blind survey a more promising prospect in the future and a testable prediction of the picture explored in this work.

\bigskip\bigskip
\noindent {\bf ACKNOWLEDGEMENTS}

\medskip

The work was funded in part by an NSF grant AST-2009619 (PK), a NASA grant 80NSSC24K0770 (PK and PB), a grant (no. 2020747) from the United States-Israel Binational Science Foundation (BSF), Jerusalem, Israel (PB) and by a grant (no. 1649/23) from the Israel Science Foundation (PB).



\end{document}